\documentclass[amsmath,amssymb,12pt]{article}

 \usepackage{color}

\usepackage{graphicx}
\usepackage{amsfonts,amsmath,amssymb}
\usepackage{mathrsfs}

\newtheorem{theorem}            {Theorem}[section]

\newtheorem{definition}         [theorem]{Definition}

\newtheorem{sideremark}         [theorem]{Remark}
\newtheorem{sideeg}           [theorem]{Example}
\newtheorem{sideconj}           [theorem]{Conjecture}
\newtheorem{sideassumption}   [theorem]{Assumption}

\newenvironment{remark}         {\begin{sideremark}\rm}{\end{sideremark}}

\usepackage{enumerate,cite,latexsym,graphicx}

\begin{document}

\title{The Series Product and Its Application to Quantum Feedforward and Feedback Networks}
\author{John~Gough\thanks{J.~Gough is with the Institute for Mathematical and Physical Sciences, University of Wales,
Aberystwyth, Ceredigion, SY23 3BZ, Wales}
 \and
Matthew R.~James\thanks{M.R. James is with the Department of
    Engineering, Australian
    National University, Canberra, ACT 0200,
    Australia. Matthew.James@anu.edu.au. Research supported by the
    Australian Research Council.}
     }

\date{\today}

\maketitle


\begin{abstract}
The purpose of this paper is to present simple and general
algebraic methods for describing series connections in quantum
networks. These methods build on and generalize existing methods
for series (or cascade) connections by allowing for more general
interfaces, and by introducing an efficient algebraic tool, the
series product. We also introduce another product, which we call
the concatenation product, that is   useful for assembling and
representing systems without necessarily having connections. We
show how the concatenation and series products can be used to
describe feedforward and feedback networks.  A selection of
examples   from the quantum control  literature are analyzed to
illustrate the utility of our network modeling methodology.

 {\bf Keywords:}   Quantum control, quantum networks, series, cascade, feedforward, feedback,  quantum noise.
\end{abstract}


\section{Introduction}
\label{sec:introduction}

\begin{figure}[h]
\begin{center}
\setlength{\unitlength}{1968sp}   
\begingroup\makeatletter\ifx\SetFigFont\undefined%
\gdef\SetFigFont#1#2#3#4#5{%
  \reset@font\fontsize{#1}{#2pt}%
  \fontfamily{#3}\fontseries{#4}\fontshape{#5}%
  \selectfont}%
\fi\endgroup%
\begin{picture}(7544,2780)(2379,-5969)
\put(6001,-4711){\makebox(0,0)[lb]{\smash{{\SetFigFont{7}{8.4}{\familydefault}{\mddefault}{\updefault}{\color[rgb]{0,0,0}$u_2=y_1$}%
}}}}
\thicklines
{\color[rgb]{0,0,0}\put(2401,-4411){\vector( 1, 0){1500}}
}%
{\color[rgb]{0,0,0}\put(6901,-5161){\framebox(1500,1500){}}
}%
{\color[rgb]{0,0,0}\put(5401,-4411){\vector( 1, 0){1500}}
}%
{\color[rgb]{0,0,0}\put(8401,-4411){\vector( 1, 0){1500}}
}%
{\color[rgb]{0,0,0}\put(3301,-5611){\dashbox{150}(5700,2400){}}
}%
\put(4501,-4486){\makebox(0,0)[lb]{\smash{{\SetFigFont{7}{8.4}{\familydefault}{\mddefault}{\updefault}{\color[rgb]{0,0,0}$\mathcal{G}_1$}%
}}}}
\put(7501,-4486){\makebox(0,0)[lb]{\smash{{\SetFigFont{7}{8.4}{\familydefault}{\mddefault}{\updefault}{\color[rgb]{0,0,0}$\mathcal{G}_2$}%
}}}}
\put(5851,-5911){\makebox(0,0)[lb]{\smash{{\SetFigFont{7}{8.4}{\familydefault}{\mddefault}{\updefault}{\color[rgb]{0,0,0}$\mathcal{G}$}%
}}}}
\put(2401,-4711){\makebox(0,0)[lb]{\smash{{\SetFigFont{7}{8.4}{\familydefault}{\mddefault}{\updefault}{\color[rgb]{0,0,0}$u_1$}%
}}}}
\put(9451,-4711){\makebox(0,0)[lb]{\smash{{\SetFigFont{7}{8.4}{\familydefault}{\mddefault}{\updefault}{\color[rgb]{0,0,0}$y_2$}%
}}}}
{\color[rgb]{0,0,0}\put(3901,-5161){\framebox(1500,1500){}}
}%
\end{picture}%

\caption{Series connection of two (classical, or non-quantum)  linear systems, denoted $\mathcal{G}=\mathcal{G}_2 \triangleleft \mathcal{G}_1$.}
\label{fig:cl-2}
\end{center}
\end{figure}
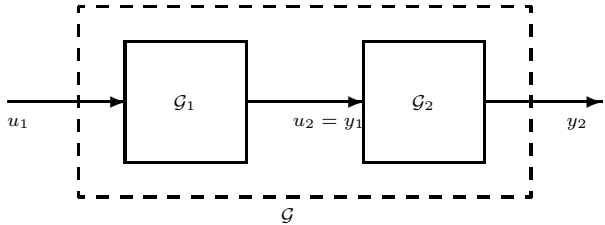

Engineers routinely use a wide range of methods and tools to help
them analyze and design control systems. For instance, control
engineers often use block diagrams to represent feedforward and
feedback systems, Figure \ref{fig:cl-2}. Among the methods that
have been developed to assist engineers are those concerning the
connection of components or subsystems to form a network. One of
the most basic  connections is the series connection, where the
output of one component is fed into the input of another, as in
Figure \ref{fig:cl-2}. When the components are (classical, or
non-quantum) linear systems, the connected system can be described
in the frequency domain by a transfer function
$\mathcal{G}(s)=\mathcal{G}_2(s) \mathcal{G}_1(s)$ which is the
product of the transfer functions of the components. The
description can also be expressed in the time domain in terms of
the state space parameters $\mathcal{G}=(A,B,C,D)$  (as we briefly
review in section \ref{sec:prelim-classical-linear}).   The series
connection  has an algebraic character, and can be regarded as a
product, $\mathcal{G}=\mathcal{G}_2 \triangleleft \mathcal{G}_1$.
Because of  new imperatives concerning  quantum network analysis
and design, in particular, quantum feedback control, \cite{HW94a},
\cite{WM94b}, \cite{SL00}, \cite{WWM02}, \cite{YK03a},
\cite{DJ06}, \cite{JNP07}, \cite{G08} the purpose of this paper is
to present simple and general algebraic methods for describing
series connections in quantum networks.

\begin{figure}[h]
\begin{center}
\begin{picture}(0,0)%
\includegraphics{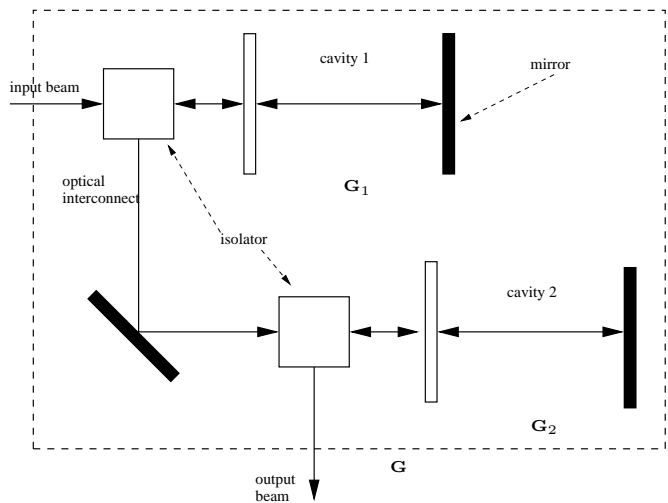}%
\end{picture}%
\setlength{\unitlength}{1934sp}%
\begingroup\makeatletter\ifx\SetFigFont\undefined%
\gdef\SetFigFont#1#2#3#4#5{%
  \reset@font\fontsize{#1}{#2pt}%
  \fontfamily{#3}\fontseries{#4}\fontshape{#5}%
  \selectfont}%
\fi\endgroup%
\begin{picture}(8444,6344)(579,-8483)
\put(4876,-4486){\makebox(0,0)[lb]{\smash{{\SetFigFont{6}{7.2}{\familydefault}{\mddefault}{\updefault}{\color[rgb]{0,0,0}$\mathbf{G}_1$}%
}}}}
\put(7276,-7561){\makebox(0,0)[lb]{\smash{{\SetFigFont{6}{7.2}{\familydefault}{\mddefault}{\updefault}{\color[rgb]{0,0,0}$\mathbf{G}_2$}%
}}}}
\put(5476,-8086){\makebox(0,0)[lb]{\smash{{\SetFigFont{6}{7.2}{\familydefault}{\mddefault}{\updefault}{\color[rgb]{0,0,0}$\mathbf{G}$}%
}}}}
\end{picture}%

\caption{Series connection of two optical cavities  via an optical interconnect (light beam) or quantum \lq\lq{wire}\rq\rq,  denoted $\mathbf{G}=\mathbf{G}_2 \triangleleft \mathbf{G}_1$.
Each cavity consists of a pair of mirrors, one of which is perfectly reflecting (shown solid)
while the other is  partially transmitting (shown unfilled). The partially transmitting mirror enables the  light mode inside the cavity to interact with an external light field, such as a laser beam. The external field is separated into input and output components by a Faraday isolator. The optical interconnect is formed when light from the output of one cavity is directed  into the input of the other, here using an additional mirror.
}
\label{fig:cavity-series}
\end{center}
\end{figure}

The types of quantum networks we consider include those arising in quantum optics, such as the optical network shown in Figure \ref{fig:cavity-series}. This network consists of a pair of optical cavities  (discussed  in subsection \ref{sec:prelim-cavity}) connected in series by a light beam which serves as  an optical interconnect or quantum \lq\lq{wire}\rq\rq.
  In this paper (section \ref{sec:series-apps}) we show how   series connections of quantum components such as this may be described as a {\em series product} $\mathbf{G}=\mathbf{G}_2 \triangleleft \mathbf{G}_1$. This product is defined in terms of system parameters $\mathbf{G}=(\mathbf{S}, \mathbf{L}, H)$, where $H$ specifies the internal energy of the system, and  $\mathcal{I}=(\mathbf{S}, \mathbf{L})$ specifies the interface of the system to external field channels (as explained in subsection \ref{prelim:open} and section \ref{sec:open-qsm}).

Series (also called cascade) connections of quantum optical
components were first considered in the papers \cite{CWG93},
\cite{HJC93}, and   certain linear feedback networks were
considered in \cite{YK03a}. Our results extend the series
connection results in these works by including more general
interfaces, and by introducing an efficient algebraic tool, the
series product. We also introduce another product, which we call
the {\em concatenation product} $\mathbf{G}=\mathbf{G}_1 \boxplus
\mathbf{G}_2$, that is  useful for assembling and representing
systems without necessarily having connections. Both products may
be used to describe a wide range of open quantum physical systems
(including those with physical variables that evolve nonlinearly)
and networks of such systems (with boson field interconnects such
as optical beams or phonon  vibrations in materials).
 We believe our modeling framework is
 of fundamental system-theoretic interest. The need for general and
efficient methods for describing networks of quantum components has been recognized to
some extent and has begun to emerge in the quantum optics and
quantum information and computing literature, e.g. \cite{YD84},
\cite{CWG93}, \cite{HJC93}, \cite[Chapter 12]{GZ00}, \cite[Chapter
4]{NC00}, \cite{YK03a},  \cite{DJ06}. It is expected  that   an
effective  quantum network theory will assist the design of
quantum technologies, just as electrical network theory and block diagram manipulations help
engineers design filters, control systems, and many other
classical electrical systems.

Series connections provide the foundation for some important
developments in quantum feedback control, e.g. \cite{HW94a},
\cite{WM94b}, \cite{WWM02}, \cite{YK03a}, \cite{JNP07},
\cite{G08}, \cite{GGY08}. To illustrate the power and utility of
our quantum network modeling methodology, we analyze several
examples from this literature. The series and concatenation
products allow us to express these quantum feedback control and
quantum filtering examples in a simple, transparent way (there are
some subtle technical issues in some of the examples for which we
provide explanation and references). We hope this will help open
up some of the quantum feedback control literature to control
engineers, which at present is largely unknown outside the physics
community. A number of articles and books  are available to help
readers with the background material on which the present  paper
is based. The papers \cite{YK03a} and  \cite{HSM05} provide
excellent introductions to aspects of the quantum models we use.
The paper \cite{BHJ07} is a tutorial article written to assist
control theorists and engineers by providing introductory
discussions of quantum mechanics, open quantum stochastic models,
and quantum filtering. The book \cite{GZ00} is an invaluable
resource for quantum noise models and quantum optics, while the
book \cite{KRP92} provides a detailed mathematical treatment of
the Hudson-Parthasarathy theory of the quantum stochastic
calculus. The book \cite{EM98} is a standard textbook on quantum
mechanics.

We begin in section \ref{sec:prelim-classical-linear} by
discussing an analog of our results in the context of classical
linear systems theory, elaborating further on the discussion at
the beginning of this section. In  section \ref{sec:prelim} we
provide a review of  some example quantum components (including
the cavity mentioned above) and connections. This section includes
a brief discussion of quantum mechanics, introduces  examples of
parametric representations, and  provides a glimpse of how the
general theory can be used. Open quantum stochastic models are
described in more detail in section \ref{sec:open-qsm}. The main
definitions and results concerning the concatenation and series
products are given in section \ref{sec:series-apps}; in
particular,   the {\em principle  of series connections}, Theorem
\ref{thm:series-fb}. In general the series product is not
commutative, but we are able to show how the order can be
interchanged by modifying one of the components, Theorem
\ref{thm:exchange}. A selection of examples   from the quantum
control  literature are analyzed    in section \ref{sec:eg}.
 The appendices contain proofs of some of the results and some
additional technical material.

{\em Notation.} In this paper we use matrices $\mathbf{M} = \{
m_{ij} \}$ with entries $m_{ij}$ that are operators on an
underlying Hilbert space. The asterisk $\ast$ is used to indicate
the Hilbert space adjoint $A^\ast$ of an operator $A$, as well as
the complex conjugate $z^\ast = x-iy$ of a complex number $z=x+i
y$ (here, $i=\sqrt{-1}$ and $x,y$ are real).  Real and imaginary
parts are denoted $\mathrm{Re}(z)=(z+z^\ast)/2$ and
$\mathrm{Im}(z)=-i(z-z^\ast)/2$ respectively. The conjugate
transpose $\mathbf{M}^\dagger$ of a matrix $\mathbf{M}$ is defined
by $\mathbf{M}^\dagger = \{ m_{ji}^\ast \}$. Also defined are the
conjugate $\mathbf{M}^\sharp = \{ m_{ij}^\ast \}$ and transpose
$\mathbf{M}^T = \{ m_{ji} \}$ matrices, so that
$\mathbf{M}^\dagger=(\mathbf{M}^T)^\sharp=(\mathbf{M}^\sharp)^T$.
In the physics literature, it is common to use the dagger
$\dagger$ to indicate the Hilbert space adjoint. The commutator of
two operators $A,B$ is defined by $[A,B]=AB-BA$. $\delta(\cdot)$
is the Dirac delta function, and $\delta_{jk}$ is the Kronecker
delta. The tensor product  of operators $A$, $B$ defined on
Hilbert spaces $\mathsf{H}$, $\mathsf{G}$ is an operator $A
\otimes B$ defined on the Hilbert space $\mathsf{H} \otimes
\mathsf{G}$ (tensor product of Hilbert spaces) defined by $(A
\otimes B) (\psi \otimes \phi) = (A\psi)  \otimes (B\phi)$ for
$\psi \in \mathsf{H}$, $\phi \in \mathsf{G}$; we usually follow
the standard shorthand and write simply $AB = A\otimes B$ for the
tensor product, and also $A=A\otimes I$ and $B=I \otimes B$.

\section{Classical Linear Systems}
\label{sec:prelim-classical-linear}

As mentioned in the Introduction (section \ref{sec:introduction}),
it is common practice in classical linear control theory  to
perform manipulations of block diagrams. Such manipulations, of
course,  greatly assist the analysis and design of control
systems. To assist readers in interpreting the main quantum
results concerning series and concatenation products  (section
\ref{sec:series-apps}), we describe concatenation and series
products for familiar classical linear systems in algebraic terms.

Consider two classical deterministic linear state space models
\begin{eqnarray}
\dot x_j &=& A_j x_j + B_j u_j
\nonumber \\
y_j &=& C_j x_j + D_j u_j
\label{cl-1}
\end{eqnarray}
where $j=1,2$. As usual, $x_j$, $u_j$ and $y_j$ are vectors and
$A_j$, $B_j$, $C_j$ and $D_j$ are appropriately sized matrices.
These systems are often represented by the matrix
\begin{equation}
\mathcal{G}_j = \left(   \begin{array}{cc} A_j & B_j
\\ C_j & D_j
\end{array} \right) ,
\label{cl-2}
\end{equation}
or the transfer function $\mathcal{G}_j(s) = C_j
(sI-A_j)^{-1}B_j+D_j$.

\begin{figure}[h]
\begin{center}

\setlength{\unitlength}{2068sp}
\begingroup\makeatletter\ifx\SetFigFont\undefined%
\gdef\SetFigFont#1#2#3#4#5{%
  \reset@font\fontsize{#1}{#2pt}%
  \fontfamily{#3}\fontseries{#4}\fontshape{#5}%
  \selectfont}%
\fi\endgroup%
\begin{picture}(4608,4880)(2379,-8069)
\put(2401,-4711){\makebox(0,0)[lb]{\smash{{\SetFigFont{7}{8.4}{\familydefault}{\mddefault}{\updefault}{\color[rgb]{0,0,0}$u_1$}%
}}}}
\thicklines
{\color[rgb]{0,0,0}\put(3901,-7261){\framebox(1500,1500){}}
}%
{\color[rgb]{0,0,0}\put(2401,-4411){\vector( 1, 0){1500}}
}%
{\color[rgb]{0,0,0}\put(2401,-6511){\vector( 1, 0){1500}}
}%
{\color[rgb]{0,0,0}\put(5401,-6511){\vector( 1, 0){1500}}
}%
{\color[rgb]{0,0,0}\put(5401,-4411){\vector( 1, 0){1500}}
}%
{\color[rgb]{0,0,0}\put(3151,-7711){\dashbox{150}(2925,4500){}}
}%
\put(4501,-6586){\makebox(0,0)[lb]{\smash{{\SetFigFont{7}{8.4}{\familydefault}{\mddefault}{\updefault}{\color[rgb]{0,0,0}$\mathcal{G}_2$}%
}}}}
\put(4501,-4486){\makebox(0,0)[lb]{\smash{{\SetFigFont{7}{8.4}{\familydefault}{\mddefault}{\updefault}{\color[rgb]{0,0,0}$\mathcal{G}_1$}%
}}}}
\put(4426,-8011){\makebox(0,0)[lb]{\smash{{\SetFigFont{7}{8.4}{\familydefault}{\mddefault}{\updefault}{\color[rgb]{0,0,0}$\mathcal{G}$}%
}}}}
\put(6451,-6811){\makebox(0,0)[lb]{\smash{{\SetFigFont{7}{8.4}{\familydefault}{\mddefault}{\updefault}{\color[rgb]{0,0,0}$y_2$}%
}}}}
\put(6526,-4786){\makebox(0,0)[lb]{\smash{{\SetFigFont{7}{8.4}{\familydefault}{\mddefault}{\updefault}{\color[rgb]{0,0,0}$y_1$}%
}}}}
\put(2401,-6811){\makebox(0,0)[lb]{\smash{{\SetFigFont{7}{8.4}{\familydefault}{\mddefault}{\updefault}{\color[rgb]{0,0,0}$u_2$}%
}}}}
{\color[rgb]{0,0,0}\put(3901,-5161){\framebox(1500,1500){}}
}%
\end{picture}%

\caption{Concatenation product.}
\label{fig:cl-1}
\end{center}
\end{figure}
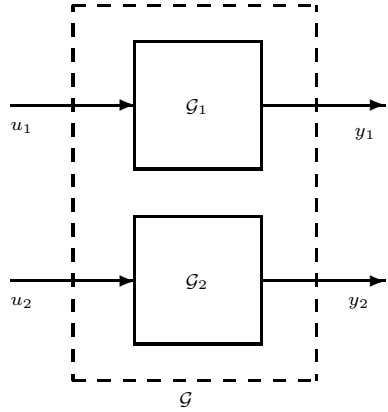

In modeling networks of such systems, one may form the {\em concatenation  product}
\begin{equation}
\mathcal{G} = \mathcal{G}_1 \boxplus \mathcal{G}_2 \nonumber  =
 \left(   \begin{array}{cc}
 \left(   \begin{array}{cc}
A_1 & 0
\\ 0 & A_2
\end{array} \right)  &  \left(   \begin{array}{cc}
B_1 &0
\\0 & B_2
\end{array} \right) \vspace{1mm}
\\  \left(   \begin{array}{cc}
C_1 & 0
\\ 0 & C_2
\end{array} \right) &  \left(   \begin{array}{cc}
D_1 & 0
\\ 0 & D_2
\end{array} \right)
\end{array} \right) ,
\label{cl-3}
\end{equation}
see Figure \ref{fig:cl-1}. In terms of transfer functions, the concatenation of two systems is $\mathcal{G}(s)  = \mathrm{diag} \{ \mathcal{G}_1(s) , \mathcal{G}_2(s)\}$. 
The concatenation  product simply assembles the two components
together, without making any connections between them. It is {\em
not} a parallel connection.

 Of considerable importance is the series connection, described by
  {\em
series product}
\begin{equation}
\mathcal{G} = \mathcal{G}_2 \triangleleft  \mathcal{G}_1 \nonumber
=
 \left(   \begin{array}{cc}
 \left(   \begin{array}{cc}
A_1 & 0
\\ B_2C_1  & A_2
\end{array} \right)  &  \left(   \begin{array}{c}
B_1
\\ B_2D_1
\end{array} \right)
\vspace{1mm} \\
 \left(   \begin{array}{cc}
D_2C_1  & C_2
\end{array} \right) & D_2D_1
\end{array} \right) ,
\label{cl-4}
\end{equation}
see Figure \ref{fig:cl-2}. Here the connection is specified by $u_2=y_1$, and so we require dim$\,u_2=$dim$\,y_1$.
In the frequency domain, the series product is given by the matrix transfer function product 
$\mathcal{G}(s) = \mathcal{G}_2(s) \mathcal{G}_1(s) 
$.
This product describes a series (or cascade) connection which is fundamental to feedforward and feedback control.

Notice that both products are defined in terms of system
parameters (state space parameters or transfer function matrices).

\section{Example Components and Connections}
\label{sec:prelim}

\subsection{Some Introductory Quantum Mechanics}
\label{sec:prelim-qm}

Central to quantum mechanics are the notions of observables $X$,
which are mathematical representations of physical quantities that
can (in principle) be measured, and state vectors $\psi$, which
summarize the status of physical systems and permit the
calculation of expectations of observables. State vectors may be
described mathematically as elements of a Hilbert space
$\mathsf{H}$, while observables are self-adjoint operators on
$\mathsf{H}$.  The expected value of an observable $X$ when in
state $\psi$ is given by the inner product  $\langle \psi, X \psi
\rangle$.

A basic example is that of a particle moving in a potential well,
\cite[Chapter 14]{EM98}. The position and momentum of the particle
are represented by   observables $Q$ and $P$, respectively,
defined by
$$
(Q \psi)(q) = q \psi (q), \ \ \ (P\psi)(q) = -i \hbar \frac{d}{dq} \psi(q)
$$
for $\psi \in \mathsf{H} = L^2(\mathbf{R})$. Here, $i=\sqrt{-1}$,
$\hbar = h/2\pi$, $h$ is Planck's constant, and $q\in \mathbf{R}$
represents  position values. In   following subsections we use
units such that $\hbar=1$, but retain it in our expressions in
this subsection. The position and momentum operators satisfy the
commutation relation $[Q,P]=i \hbar$. The dynamics of the particle
is given by {\em Schr\"{o}dinger's equation}
$$
i \hbar \frac{d}{dt} V(t) = H V(t) ,
$$
with initial condition $V(0)=I$,
where
$
H = \frac{P^2}{2m} + \frac{1}{2}  m \omega^2 Q^2
$
is the {\em Hamiltonian} (here, $m$ is the mass of the particle,
and $\omega$ is the frequency of oscillation). The operator $V(t)$
is unitary ($V^\ast(t) V(t)= V(t) V^\ast(t) = I$, where $I$ is the
identity operator, and the asterisk denotes Hilbert space
adjoint)---it is analogous to the transition matrix in classical
linear systems theory. State vectors and observables evolve
according to
$$
\psi_t = V(t) \psi \in \mathsf{H},  \ \ \ X(t) = V^\ast(t) X V(t) .
$$
These expressions  provide two equivalent descriptions (dual), the
former is referred to as the {\em Schr\"{o}dinger picture}, while
the latter is the {\em Heisenberg picture}. In this paper we use
the Heisenberg picture, which is more closely related to models
used in classical control theory and classical probability theory.
In the Heisenberg picture, observables (and more generally other
operators  on $\mathsf{H}$) evolve according to
\begin{equation}
\frac{d}{dt} X(t) = -\frac{i}{\hbar} [ X(t), H(t) ] ,
\label{prelim:hp1}
\end{equation}
where $H(t)=V^\ast(t) H V(t)$.

Energy eigenvectors $\psi_n$ are defined by the equation $H \psi_n
= E_n \psi_n$ for real numbers $E_n$. The system has a discrete
energy spectrum $E_n = (n+\frac{1}{2}) \hbar \omega$,
$n=0,1,2,\ldots$. The state $\psi_0$ corresponding to $E_0$ is
called the {\em ground state}. The {\em annihilation operator}
$$
a = \sqrt{\frac{m \omega}{2\hbar} }(Q+ i \frac{P}{2m\omega})
$$
and the creation operator $a^\ast$ lower and raise energy levels,
respectively: $a \psi_n =  \sqrt{n} \psi_{n-1}$, and $a^\ast
\psi_n = \sqrt{n+1} \psi_{n+1}$. They satisfy the canonical
commutation relation $[a,a^\ast]=1$. In terms of these operators,
the Hamiltonian  can be expressed as $H = \hbar \omega( a^\ast a +
\frac{1}{2}) . $ Using (\ref{prelim:hp1}), the annihilation
operator evolves according to
\begin{equation}
\frac{d}{dt} a(t) = -i\omega a(t)
\label{ho-a}
\end{equation}
with solution $a(t)=e^{-i \omega t} a$.  Note that also $a^\ast(t)
= e^{i \omega t} a^\ast$, and so commutation relations are
preserved by the unitary dynamics:
$[a(t),a^\ast(t)]=[a,a^\ast]=1$. Because of the oscillatory nature
of the dynamics, this system is often refereed to as the {\em
quantum harmonic oscillator}.

It can be seen that the Hamiltonian $H$ is a key
\lq\lq{parameter}\rq\rq \ of the quantum physical system,
specifying  its  energy.

\subsection{Optical Cavities}
\label{sec:prelim-cavity}

\begin{figure}[h]
\begin{center}

 \begin{picture}(0,0)%
\includegraphics{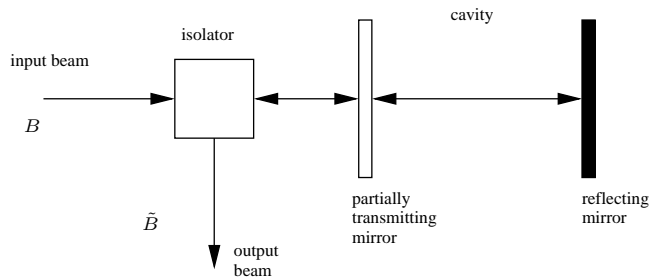}%
\end{picture}%
\setlength{\unitlength}{2171sp}%
\begingroup\makeatletter\ifx\SetFigFont\undefined%
\gdef\SetFigFont#1#2#3#4#5{%
  \reset@font\fontsize{#1}{#2pt}%
  \fontfamily{#3}\fontseries{#4}\fontshape{#5}%
  \selectfont}%
\fi\endgroup%
\begin{picture}(7385,3111)(-6689,-5401)
\put(-6524,-3736){\makebox(0,0)[lb]{\smash{{\SetFigFont{7}{8.4}{\familydefault}{\mddefault}{\updefault}{\color[rgb]{0,0,0}$B$}%
}}}}
\put(-5174,-4861){\makebox(0,0)[lb]{\smash{{\SetFigFont{7}{8.4}{\familydefault}{\mddefault}{\updefault}{\color[rgb]{0,0,0}$\tilde{B}$}%
}}}}
\end{picture}%

\caption{A  cavity consists of a pair of mirrors, one of which
is perfectly reflecting (shown solid) while the other is partially
transmitting (shown unfilled). The partially transmitting mirror
enables the  light mode inside the cavity to interact with an
external light field, such as a laser beam. The external field is
separated into input and output components by a Faraday isolator.
}
\label{fig:cavity-a}
\end{center}
\end{figure}

\begin{figure}[h]
\begin{center}

 \begin{picture}(0,0)%
\includegraphics{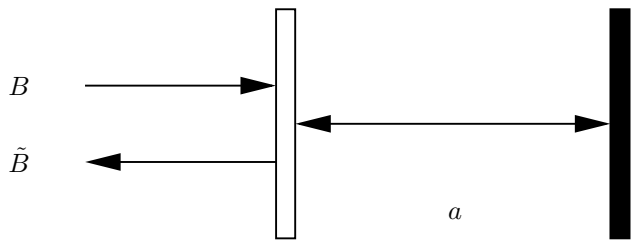}%
\end{picture}%
\setlength{\unitlength}{3158sp}%
\begingroup\makeatletter\ifx\SetFigFont\undefined%
\gdef\SetFigFont#1#2#3#4#5{%
  \reset@font\fontsize{#1}{#2pt}%
  \fontfamily{#3}\fontseries{#4}\fontshape{#5}%
  \selectfont}%
\fi\endgroup%
\begin{picture}(4912,1844)(1486,-4283)
\put(4951,-4111){\makebox(0,0)[lb]{\smash{{\SetFigFont{10}{12.0}{\familydefault}{\mddefault}{\updefault}{\color[rgb]{0,0,0}$a$}%
}}}}
\put(1501,-3136){\makebox(0,0)[lb]{\smash{{\SetFigFont{10}{12.0}{\familydefault}{\mddefault}{\updefault}{\color[rgb]{0,0,0}$B$}%
}}}}
\put(1501,-3736){\makebox(0,0)[lb]{\smash{{\SetFigFont{10}{12.0}{\familydefault}{\mddefault}{\updefault}{\color[rgb]{0,0,0}$\tilde{B}$}%
}}}}
\end{picture}%

\caption{A simplified representation of the cavity from Figure \ref{fig:cavity-a} which omits the
Faraday isolator.  It shows  input $B$ and output $\tilde B$
fields and the cavity mode annihilation operator $a$. This
representation will be used for the remainder of this paper.}
\label{fig:cavity-b}
\end{center}
\end{figure}

A  diagram  of an optical cavity is shown in Figures
\ref{fig:cavity-a}, \ref{fig:cavity-b},   together with a simplified representation. It
consists of a pair of mirrors; the left one is partially
transmitting (shown unfilled), while the right mirror is assumed
perfectly reflecting (shown solid). Between the mirrors a trapped
electromagnetic (optical) mode is set up, whose frequency depends
on the separation between the mirrors. This mode is described by a
harmonic oscillator with  annihilation operator $a$ (an operator
acting on a Hilbert space $\mathsf{H}$ (as in subsection
\ref{sec:prelim-qm}), called the initial space). The partially
transmitting mirror affords the opportunity for this mode to
interact with an external free field, represented by a quantum
stochastic process $b(t)$ (to be discussed shortly). When the
external field is in the vacuum state, energy initially inside the
cavity mode may leak out, in which case the cavity system is a
damped harmonic oscillator, \cite{GZ00}.

Quantization of a (free) electromagnetic field leads to an expression for the vector potential
$$
\mathbf{A}(x,t) = \int \kappa(\omega)[ b(\omega) e^{-i\omega t + i
\omega x/c} + b^\ast(\omega) e^{i\omega t - i
\omega x/c} ] d\omega ,
$$
for a suitable coefficients $\kappa(\omega)$, and annihilation
operators $b(\omega)$. Such a field can be considered as an
infinite collection of harmonic oscillators, satisfying the
singular canonical commutation relations
$$
[b(\omega), b^\ast(\omega') ] = \delta(\omega-\omega') ,
$$
where $\delta$ is the Dirac delta function.

An optical signal, such as a laser beam, is a free field with
frequency content concentrated at a very high frequency $\omega_0
\approx 10^{14}$ rad/sec. The fluctuations about this nominal
frequency can be considered as a quantum stochastic process
consisting of  signal plus noise, where the noise is of high
bandwidth relative to the signal. Indeed, a coherent field is a
good, approximate,  model of a laser beam, and can be considered
as the sum $b(t)=s(t)+b_0(t)$, where $s(t)$ is a signal, and
$b_0(t)$ is quantum (vacuum) noise. Such \lq\lq{signal plus
noise}\rq\rq \ models are of course common in engineering.

The cavity mode-free field system has a natural input-output
structure, where the free field is decomposed as a superposition
of right and left traveling fields. The right traveling field
component is regarded as the {\em input}, while the left traveling
component is an {\em output}, containing information about the
cavity mode after interaction. The interaction facilitated by the
partially transmitting mirror provides a boundary condition for
the fields. The two components can be separated in the laboratory
using a Faraday isolator. This leads to idealized models based on
rotating wave and Markovian approximations, where, in the time
domain, the input optical field (when in the ground or vacuum
state) is described by {\em quantum white noise} $b(t)=b_0(t)$
\cite[Chapters 5 and 11]{GZ00}, which satisfies the singular
canonical commutation relations
\begin{equation}
[b(t), b^\ast(t') ] = \delta(t-t') .
\label{qwn1}
\end{equation}
In order to accommodate such singular processes, rigorous white
noise and It\={o} frameworks have been developed, where in the
It\={o} theory one uses the integrated noise, informally written
$$
B(t) = \int_0^t b(s) ds .
$$
The operators $B(t)$  are defined on a particular Hilbert space
called a {\em Fock space}, $\mathsf{F}$, \cite[sec. 19]{KRP92}.
When the field is in the vacuum (or ground) state, this is the
quantum Wiener process which satisfies the It\={o} rule
$$
dB(t) dB^\ast(t) = dt
$$
(all other It\={o} products are zero). Field quadratures, such as
$B(t)+B^\ast(t)$ and $-i(B(t)-B^\ast(t))$ are each equivalent to
classical Wiener processes, but do not commute. A field quadrature
can be measured using homodyne detection, \cite[Chapter 8]{GZ00}.

The cavity mode-free field system can be described by the Hamiltonian
\begin{equation}
H = \Delta a^\ast a - i \hbar \int k(\omega)  ( a^\ast b(\omega) - b^\ast(\omega) a) d\omega ,
\label{cavity-H}
\end{equation}
where the first term represents the self-energy of the cavity mode
(the number $\Delta$ is called the \lq\lq{detuning}\rq\rq, and
represents the difference between the nominal external field
frequency and the cavity mode frequency), while the remaining two
terms describe the energy flow between the cavity mode and the
free field (a photon in the free field may be created by a loss of
a photon from the cavity mode, and vice versa). This Hamiltonian
is defined on the composite Hilbert space, the tensor product
$\mathsf{H} \otimes \mathsf{F}$; the tensor product is not written
explicitly in the expression (\ref{cavity-H}).

The Schr\"{o}dinger equation for the cavity-free field system is
derived from (\ref{cavity-H}) under certain assumptions
\cite{GZ00}, and is given by the It\={o} quantum stochastic
differential equation (QSDE)
\begin{eqnarray}
d V(t) = \{   \sqrt{\gamma} a dB^\ast(t) - \sqrt{\gamma} a^\ast dB(t)
\nonumber \\
- \frac{\gamma}{2} a^\ast a dt -i \Delta a^\ast a dt
\} V(t) ,
\label{prelim:hp5}
\end{eqnarray}
with vacuum input and initial condition $V(0)=I$, so that $V(t)$
is unitary. The complete cavity mode-free field system thus has a
unitary model. In the Heisenberg picture,   cavity mode operators
$X$ (operators on the initial space $\mathsf{H}$) evolve according
the quantum It\={o} equation
\begin{eqnarray}
dX(t) & =& -i \Delta [ X(t), a^\ast(t) a(t) ]dt
\label{prelim:hp3}  \\
&+& \frac{\gamma}{2} (  a^\ast (t) [X(t),a(t)]   + [a^\ast (t), X(t)] a(t)  )dt
\notag \\
&+& \sqrt{\gamma} dB^\ast(t)[X(t), a(t) ]
 + \sqrt{\gamma} [a^\ast(t), X(t) ] dB(t) .
\notag
\end{eqnarray}
Here, $\gamma > 0$ is a parameter specifying the coupling strength, and is related an approximation of the function $k(\omega)$ in the Hamiltonian (\ref{cavity-H}).
In particular, for $X=a$, the cavity mode annihilation operator, we have
\begin{equation}
d a(t) = -(\frac{\gamma}{2} +i \Delta) a(t) dt - \sqrt{\gamma}\, dB(t) ;
\label{prelim:hp4}
\end{equation}
cf.  (\ref{ho-a}).
The output field $\tilde B(t)$ is given by
\begin{equation}
d \tilde B(t) = \sqrt{\gamma}\, a(t) dt +  dB(t) ,
\label{prelim:hp4a}
\end{equation}
where one can see the \lq\lq{signal plus noise}\rq\rq \ form of the field.

This is an example of an {\em open quantum system}, characterized by the parameters
$\sqrt{\gamma} a$ and $\Delta a^\ast a$;
the latter being the cavity mode Hamiltonian (specifying internal energy), and the former being the operator coupling the cavity mode to the external field (specifying the interface). These parameters are operators defined on the initial space $\mathsf{H}$. These parameters specify a simpler, idealized model employing quantum noise, in place of the more basic but complicated Hamiltonian (\ref{cavity-H}).


\subsection{Optical Beamsplitters}
\label{sec:prelim-bs}

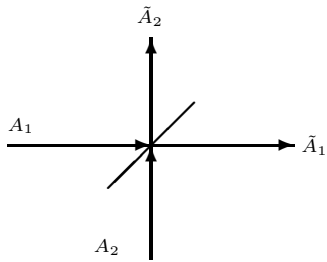
\begin{figure}[h]
\begin{center}

\setlength{\unitlength}{2368sp}%
\begingroup\makeatletter\ifx\SetFigFont\undefined%
\gdef\SetFigFont#1#2#3#4#5{%
  \reset@font\fontsize{#1}{#2pt}%
  \fontfamily{#3}\fontseries{#4}\fontshape{#5}%
  \selectfont}%
\fi\endgroup%
\begin{picture}(3097,2653)(579,-4283)
\put(3676,-3136){\makebox(0,0)[lb]{\smash{{\SetFigFont{7}{8.4}{\familydefault}{\mddefault}{\updefault}{\color[rgb]{0,0,0}$\tilde{A}_1$}%
}}}}
\thicklines
{\color[rgb]{0,0,0}\put(2551,-2611){\line(-1,-1){900}}
}%
{\color[rgb]{0,0,0}\put(601,-3061){\vector( 1, 0){1500}}
}%
{\color[rgb]{0,0,0}\put(2101,-4261){\vector( 0, 1){1200}}
}%
{\color[rgb]{0,0,0}\put(2101,-3061){\vector( 0, 1){1125}}
}%
\put(601,-2911){\makebox(0,0)[lb]{\smash{{\SetFigFont{7}{8.4}{\familydefault}{\mddefault}{\updefault}{\color[rgb]{0,0,0}$A_1$}%
}}}}
\put(1501,-4186){\makebox(0,0)[lb]{\smash{{\SetFigFont{7}{8.4}{\familydefault}{\mddefault}{\updefault}{\color[rgb]{0,0,0}$A_2$}%
}}}}
\put(1951,-1786){\makebox(0,0)[lb]{\smash{{\SetFigFont{7}{8.4}{\familydefault}{\mddefault}{\updefault}{\color[rgb]{0,0,0}$\tilde{A}_2$}%
}}}}
{\color[rgb]{0,0,0}\put(2101,-3061){\vector( 1, 0){1500}}
}%
\end{picture}%

\caption{Diagram of an optical beamsplitter showing inputs $A_1, A_2$ and outputs  $\tilde A_1, \tilde A_2$ fields.}
\label{fig:bs}
\end{center}
\end{figure}

A beamsplitter is a device that effects the interference of incoming optical fields $A_1, A_2$ and produces outgoing optical fields $\tilde A_1, \tilde A_2$, Figure \ref{fig:bs}. The relationship between these fields is
\begin{equation}
\tilde{A}_1(t) = \beta A_1(t) - \alpha A_2(t) , \quad
\tilde{A}_2(t) = \alpha A_1(t) +\beta  A_2(t) ,
\label{prelim:bs-1}
\end{equation}
where $\alpha$ and $\beta$ are complex numbers describing the
beamsplitter relations, and they satisfy $\alpha^\ast \alpha +
\beta^\ast \beta =1$, $\alpha^\ast \beta = \alpha\beta^\ast$ (here the asterisk indicates the conjugate of a
complex number).

 The initial space is trivial, $\mathsf{H}=\mathbf{C}$, the complex numbers; nevertheless, the
 Schr\"{o}dinger equation for the beamsplitter  is
\begin{equation}
d V(t) = \{    (\mathbf{S} - \mathbf{I} ) d \mathbf{\Lambda}
\} V(t) ,
\label{prelim:hp50}
\end{equation}
with initial condition $V(0)=I$,
where $\mathbf{S}$ is the unitary matrix defined by (\ref{prelim:bs-2}) below, $\mathbf{I}$ is the identity matrix, and $\mathbf{\Lambda}$ is the matrix of gauge processes
\begin{equation}
 \mathbf{\Lambda}  = \left(  \begin{array}{cc}
 A_{11} & A_{12}
 \\
 A_{21} &  A_{22}
 \end{array} \right) .
 \label{prelim:hp51}
\end{equation}
Here, $A_{ij}$ describes the destruction of a photon in channel $j$ and the creation of a photon in channel $i$. In terms of their formal derivatives, $A_{ij}(t) = \int_0^t a^\ast_i(s) a_j(s) ds$, where $A_i(t)=\int_0^t a_i(s)ds$.
The self-adjoint processes $A_{jj}$ are equivalent to classical Poisson processes when the channels are in coherent states (signal plus quantum noise). These counting processes may be observed by a photodetector,  \cite[Chapters 8 and 11]{GZ00}.

This open system is characterized by the unitary parameter matrix
\begin{equation}
\mathbf{S}=\left(  \begin{array}{cc}
\beta & -\alpha \\
\alpha & \beta
\end{array} \right) ,
\label{prelim:bs-2}
\end{equation}
which  describes scattering among the field channels. The matrix $\mathbf{S}$  specifies the interface for the beamsplitter.

\subsection{Open Quantum Systems}
\label{prelim:open}

In general, as we shall explain in more detail in section \ref{sec:open-qsm}, open quantum systems with multiple field channels are characterized by the parameter list
\begin{equation}
\mathbf{G} = (\mathbf{S}, \mathbf{L}, H)
\label{prelim-G}
\end{equation}
where $\mathbf{S}$ is a square matrix with operator entries such that $\mathbf{S}^\dagger \mathbf{S}= \mathbf{S} \mathbf{S}^\dagger = \mathbf{I}$ (recall the notational conventions mentioned at the end of section \ref{sec:introduction}), $\mathbf{L}$ is a column vector with operator entries, and $H$ is a self-adjoint operator. The matrix $\mathbf{S}$ is called a {\em scattering matrix}, the vector $\mathbf{L}$ is a {\em coupling vector}; together, these parameters specify the {\em interface} between the system and the fields. The parameter
 $H$ is the Hamiltonian describing the self-energy of the system.
 Thus the parameters describe the system by specifying energies---internal energy, and energy exchanged with the fields.
 All operators in the parameter list are defined on the initial Hilbert space $\mathsf{H}$ for the system.

The closed, undamped, harmonic oscillator of subsection \ref{sec:prelim-qm} is specified by the parameters
\begin{equation}
\mathbf{H} = (\_, \_, \omega a^\ast a)
\label{prelim-H}
\end{equation}
(the blanks $\_$ indicate the absence of field channels), while the open, damped oscillator (cavity) of  subsection \ref{sec:prelim-cavity} has parameters
 \begin{equation}
\mathbf{C} = (I, \sqrt{\gamma}\, a, \Delta a^\ast a) .
\label{prelim-C}
\end{equation}
The beamsplitter, described in subsection \ref{sec:prelim-bs} has parameters
\begin{equation}
\mathbf{M} = ( \left(  \begin{array}{cc}
\beta & -\alpha \\
\alpha & \beta
\end{array} \right) , 0, 0) .
\label{prelim-B}
\end{equation}

\subsection{Series Connection Example}
\label{sec:prelim-series}

 \begin{figure}[h]
\begin{center}

 \begin{picture}(0,0)%
\includegraphics{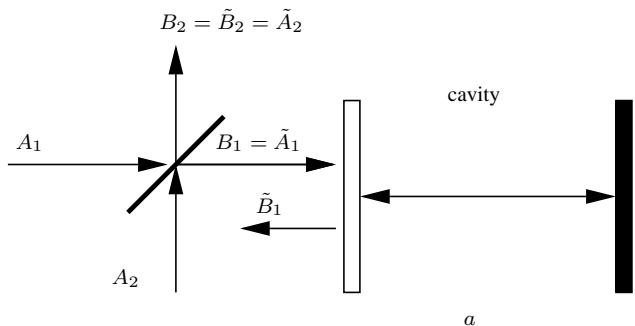}%
\end{picture}%
\setlength{\unitlength}{2644sp}%
\begingroup\makeatletter\ifx\SetFigFont\undefined%
\gdef\SetFigFont#1#2#3#4#5{%
  \reset@font\fontsize{#1}{#2pt}%
  \fontfamily{#3}\fontseries{#4}\fontshape{#5}%
  \selectfont}%
\fi\endgroup%
\begin{picture}(5894,3027)(504,-4630)
\put(601,-2911){\makebox(0,0)[lb]{\smash{{\SetFigFont{8}{9.6}{\familydefault}{\mddefault}{\updefault}{\color[rgb]{0,0,0}$A_1$}%
}}}}
\put(1501,-4186){\makebox(0,0)[lb]{\smash{{\SetFigFont{8}{9.6}{\familydefault}{\mddefault}{\updefault}{\color[rgb]{0,0,0}$A_2$}%
}}}}
\put(1951,-1786){\makebox(0,0)[lb]{\smash{{\SetFigFont{8}{9.6}{\familydefault}{\mddefault}{\updefault}{\color[rgb]{0,0,0}$B_2=\tilde{B}_2=\tilde{A}_2$}%
}}}}
\put(2851,-3511){\makebox(0,0)[lb]{\smash{{\SetFigFont{8}{9.6}{\familydefault}{\mddefault}{\updefault}{\color[rgb]{0,0,0}$\tilde{B}_1$}%
}}}}
\put(2476,-2911){\makebox(0,0)[lb]{\smash{{\SetFigFont{8}{9.6}{\familydefault}{\mddefault}{\updefault}{\color[rgb]{0,0,0}$B_1=\tilde{A}_1$}%
}}}}
\put(4801,-4561){\makebox(0,0)[lb]{\smash{{\SetFigFont{8}{9.6}{\familydefault}{\mddefault}{\updefault}{\color[rgb]{0,0,0}$a$}%
}}}}
\end{picture}%

\caption{Beam splitter (left) and cavity (right)  network.}
\label{fig:cavity-bs-1}
\end{center}
\end{figure}

Consider the feedforward  network shown in Figure \ref{fig:cavity-bs-1}, where  one of the beamsplitter output beams is fed into an optical cavity.
 From the previous subsections, we see that the  quantum
stochastic differential equations describing the network are
\begin{eqnarray}
d a(t) &=&   (- \frac{\gamma}{2}  +i\Delta )a(t) dt -\sqrt{\gamma} \, d B_1(t)
\label{cavity-1}  \\
\tilde{A}_1(t) &=& \beta A_1(t) - \alpha A_2(t)
\label{bs-1} \\
\tilde{A}_2(t) &=& \alpha A_1(t) +\beta  A_2(t)
\label{bs-2} \\
B_1(t) &=& \tilde{A}_1(t)
\label{bs-1a} \\
B_2(t) &=& \tilde{A}_2(t)
\label{bs-2a} \\
d \tilde{B}_1(t) &=& \sqrt{\gamma} a(t) dt + dB_1(t)
\label{cavity-2} \\
d \tilde{B}_2(t) &=& d B_2(t) .
\label{cavity-3}
\end{eqnarray}
 It can be seen that  algebraic manipulations are required to describe
the complete system (in general such manipulations may be simple
in principle, but complicated in practice).  The key motivation for this paper is more efficient algebraic methods for describing such networks.

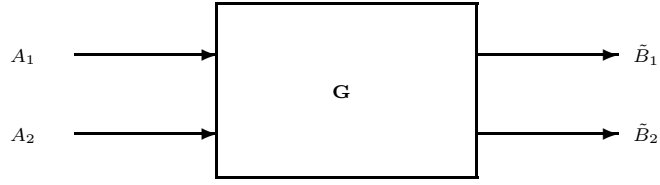
\begin{figure}[h]
\begin{center}

 \setlength{\unitlength}{2368sp}%
\begingroup\makeatletter\ifx\SetFigFont\undefined%
\gdef\SetFigFont#1#2#3#4#5{%
  \reset@font\fontsize{#1}{#2pt}%
  \fontfamily{#3}\fontseries{#4}\fontshape{#5}%
  \selectfont}%
\fi\endgroup%
\begin{picture}(6555,1844)(1411,-6683)
\put(4801,-5836){\makebox(0,0)[lb]{\smash{{\SetFigFont{7}{8.4}{\familydefault}{\mddefault}{\updefault}{\color[rgb]{0,0,0}$\mathbf{G}$}%
}}}}
\thicklines
{\color[rgb]{0,0,0}\put(2101,-5386){\vector( 1, 0){1500}}
}%
{\color[rgb]{0,0,0}\put(2101,-6211){\vector( 1, 0){1500}}
}%
{\color[rgb]{0,0,0}\put(6301,-6211){\vector( 1, 0){1500}}
}%
{\color[rgb]{0,0,0}\put(6301,-5386){\vector( 1, 0){1500}}
}%
\put(1426,-5461){\makebox(0,0)[lb]{\smash{{\SetFigFont{7}{8.4}{\familydefault}{\mddefault}{\updefault}{\color[rgb]{0,0,0}$A_1$}%
}}}}
\put(1426,-6286){\makebox(0,0)[lb]{\smash{{\SetFigFont{7}{8.4}{\familydefault}{\mddefault}{\updefault}{\color[rgb]{0,0,0}$A_2$}%
}}}}
\put(7951,-5461){\makebox(0,0)[lb]{\smash{{\SetFigFont{7}{8.4}{\familydefault}{\mddefault}{\updefault}{\color[rgb]{0,0,0}$\tilde{B}_1$}%
}}}}
\put(7951,-6286){\makebox(0,0)[lb]{\smash{{\SetFigFont{7}{8.4}{\familydefault}{\mddefault}{\updefault}{\color[rgb]{0,0,0}$\tilde{B}_2$}%
}}}}
{\color[rgb]{0,0,0}\put(3601,-6661){\framebox(2700,1800){}}
}%
\end{picture}%

\caption{Beam splitter-cavity network representation illustrating the network model given by (\ref{cavity-4}).}
\label{fig:cavity-bs-1a}
\end{center}
\end{figure}

We now describe how the parameters for the complete network may be obtained. We first assemble the field channels into vectors as follows:
\begin{eqnarray*}
\mathbf{A} = \left(   \begin{array}{c}
A_1 \\ A_2
\end{array} \right) ,   \mathbf{B} = \left(   \begin{array}{c}
 B_1 \\   B_2
\end{array} \right), 
 \tilde{\mathbf{A}} = \left(   \begin{array}{c}
 \tilde{A}_1 \\   \tilde{A}_2
\end{array} \right), 
 \tilde{\mathbf{B}} = \left(   \begin{array}{c}
 \tilde{B}_1 \\   \tilde{B}_2
\end{array} \right) .
\end{eqnarray*}
The beamsplitter acts on the input vector $\mathbf{A}$, and is
described by the parameters $\mathbf{M}$ given in equation
(\ref{prelim-B})). Now the beamsplitter output has two channels,
while the cavity has one channel (described by the parameters
$\mathbf{C}$, equation (\ref{prelim-C})), and so we augment the
cavity to accept  a second channel in a trivial way. This is
achieved by forming the concatenation $\mathbf{C} \boxplus
\mathbf{N}$, where $ \mathbf{N} =  (1, 0, 0) $ represents a
trivial component (pass-through). The augmented cavity $\mathbf{C}
\boxplus \mathbf{N}$ can now accept the output of the
beamsplitter, so that the complete network is described as a
series connection as follows:
\begin{equation}
 \mathbf{G} = ( \mathbf{C} \boxplus \mathbf{N})   \triangleleft \mathbf{M} .
\label{cavity-6}
\end{equation}
The definition of the concatenation $\boxplus$ and series
$\triangleleft$ products will be explained below in  section
\ref{sec:series-apps}  (Definitions \ref{dfn:concat} and
\ref{dfn:series}, and the principle of series connections, Theorem
\ref{thm:series-fb}).
 By applying these definitions, we obtain the network parameters
\begin{equation}
\mathbf{G} = \left(
\left(   \begin{array}{cc}
 \beta & -\alpha \\ \alpha & \beta
\end{array} \right) ,
\left(   \begin{array}{c}
 \sqrt{\gamma}\, a \\ 0
\end{array} \right) ,
\Delta a^\ast a
\right)  .
\label{cavity-4}
\end{equation}
A schematic representation of the network is shown in Figure
\ref{fig:cavity-bs-1a}, which illustrates the important point that
{\em  components, parts of components, as well as the complete network, are
described by parameters of the form (\ref{prelim-G}).}

\begin{figure}[h]
\begin{center}

 \begin{picture}(0,0)%
\includegraphics{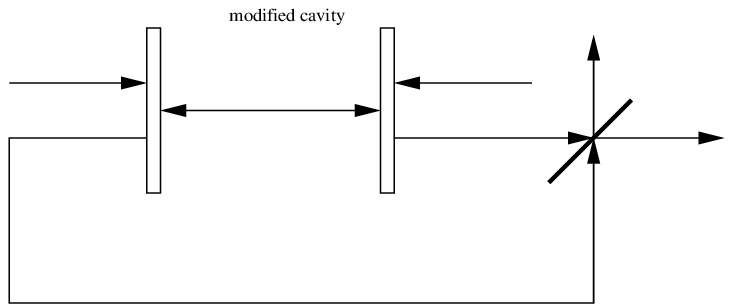}%
\end{picture}%
\setlength{\unitlength}{1737sp}%
\begingroup\makeatletter\ifx\SetFigFont\undefined%
\gdef\SetFigFont#1#2#3#4#5{%
  \reset@font\fontsize{#1}{#2pt}%
  \fontfamily{#3}\fontseries{#4}\fontshape{#5}%
  \selectfont}%
\fi\endgroup%
\begin{picture}(8587,3280)(1336,-5483)
\put(2476,-2911){\makebox(0,0)[lb]{\smash{{\SetFigFont{5}{6.0}{\familydefault}{\mddefault}{\updefault}{\color[rgb]{0,0,0}$B_1=\tilde{A}_1$}%
}}}}
\put(1351,-3136){\makebox(0,0)[lb]{\smash{{\SetFigFont{5}{6.0}{\familydefault}{\mddefault}{\updefault}{\color[rgb]{0,0,0}$A_1$}%
}}}}
\put(7201,-2911){\makebox(0,0)[lb]{\smash{{\SetFigFont{5}{6.0}{\familydefault}{\mddefault}{\updefault}{\color[rgb]{0,0,0}$A_2$}%
}}}}
\put(9526,-3436){\makebox(0,0)[lb]{\smash{{\SetFigFont{5}{6.0}{\familydefault}{\mddefault}{\updefault}{\color[rgb]{0,0,0}$\tilde{B}_2$}%
}}}}
\put(8101,-2386){\makebox(0,0)[lb]{\smash{{\SetFigFont{5}{6.0}{\familydefault}{\mddefault}{\updefault}{\color[rgb]{0,0,0}$\tilde{B}_1$}%
}}}}
\end{picture}%

\caption{Equivalent beam splitter   and cavity    network.}
\label{fig:cavity-bs-1b}
\end{center}
\end{figure}

For the purposes of network modeling and design, it can be useful to perform manipulations of the network to yield equivalent networks; this, of course, is common practice in classical  electrical circuit theory and control engineering. For instance, in our example we could move the beam splitter to the output, but the cavity should be modified (to have two partially transmitting mirrors) as follows (see Remark \ref{rmk:pass-thru}):
\begin{equation}
 \mathbf{G} = ( \mathbf{C} \boxplus \mathbf{N})   \triangleleft \mathbf{M}  =  \mathbf{M}  \triangleleft   ( \mathbf{C}' \boxplus \mathbf{N}')   .
\label{cavity-10}
\end{equation}
Here, the modified cavity $ \mathbf{C}' \boxplus \mathbf{N}'$ (see Figure \ref{fig:cavity-bs-1b}) is described by the subsystems
\begin{equation}
 \mathbf{C}' =  \left(  I, \beta^\ast \sqrt{\gamma} \, a, \Delta a^\ast a
  \right) , \ \
\mathbf{N}' = \left(    I, -\alpha^\ast \sqrt{\gamma}\, a, 0
\right) .
\label{cavity-11}
\end{equation}

The connections described here so far are unidirectional {\em
field mediated  connections}. Components interact indirectly via a
quantum field, which acts as a quantum \lq\lq{wire}\rq\rq. One can
also consider bidirectional {\em direct connections}, which can
be accommodated by using interaction Hamiltonian terms in the
models. Our emphasis in this paper will be on field mediated
connections, with direct  connections readily available in the
modeling framework if required. See subsection
\ref{sec:series-apps-reduce}.

\section{Open Quantum Stochastic Models}
\label{sec:open-qsm}

In this section we describe in more detail the open quantum models of the type encountered  in section \ref{sec:prelim}.  Specifically, we consider models specified by the parameters $\mathbf{G} = (\mathbf{S}, \mathbf{L}, H)$ (recall (\ref{prelim-G})), where
$$
\mathbf{S} =   \left( \begin{array}{ccc}
S_{11} & \hdots & S_{1n} \\
 \vdots & \vdots & \vdots \\
 S_{n1} & \hdots & S_{nn}
\end{array} \right), \ \ \mathbf{L} =  \left( \begin{array}{c}
L_1 \\ \vdots \\ L_n
\end{array} \right) ,
$$
are respectively a scattering matrix with operator entries satisfying $\mathbf{S}^\dagger \mathbf{S}= \mathbf{S} \mathbf{S}^\dagger = \mathbf{I}$, and coupling vector with operator entries, and $H$ is a self-adjoint operator called the Hamiltonian (this parameterization is due to Hudson-Parthasarathy, \cite{HP84}, and is closely related to a standard form of the Lindblad generator, given in (\ref{lindblad}) below). The operators constituting these parameters are assumed to be defined on an underlying Hilbert space $\mathsf{H}$, called the {\em initial space}.
These parameters specify an open quantum system coupled to $n$ field channels with corresponding gauge processes:
$$
\mathbf{A} =  \left( \begin{array}{c}
A_1 \\ \vdots \\ A_n
\end{array} \right),
\ \
\mathbf{\Lambda} = \left( \begin{array}{ccc}
A_{11} & \hdots & A_{1n} \\
 \vdots & \vdots & \vdots \\
 A_{n1} & \hdots & A_{nn}
\end{array} \right) .
$$
All differentials shall be understood in the It\={o}
sense - that is, $dX\left( t\right) \equiv X\left( t+dt\right) -X\left(
t\right)$.
We assume that these processes are {\em canonical}, meaning that we have the following non-vanishing second order It\={o} products: $%
dA_{j}\left( t\right) dA_{k}\left( t\right) ^{\ast }=\delta _{jk}dt$, $%
dA_{jk}\left( t\right) dA_{l}\left( t\right) ^{\ast }=\delta
_{kl}dA_{j}(t)^{\ast }$, $\,dA_{j}\left( t\right) dA_{kl}\left( t\right)
=\delta _{jk}dA_{l}(t)$ and $dA_{jk}\left( t\right) dA_{lm}\left( t\right)
=\delta _{kl}dA_{jm}(t)$.

If we consider the open system specified by $\mathbf{G} =
(\mathbf{S}, \mathbf{L}, H)$ with canonical inputs, the
Schr\"{o}dinger equation
\begin{eqnarray}
d V(t) &=& \{ \mathrm{tr}[(\mathbf{S}-\mathbf{I})
d\mathbf{\Lambda}] + d\mathbf{A}^\dagger \mathbf{L}
\label{open-hp-G}  \\
&-&  \mathbf{L}^\dagger \mathbf{S} d\mathbf{A} - \frac{1}{2}
\mathbf{L}^\dagger \mathbf{L} dt -i H dt  \} V(t) \equiv d
G(t)\thinspace V(t)
\notag
\end{eqnarray}
with initial condition $V(0)=I$ determines the unitary motion of
the system. Equation (\ref{open-hp-G}) serves as the definition of
the time-dependent generator $d G(t)$. Given an operator $X$
defined on the initial space $\mathsf{H}$, its Heisenberg
evolution is defined by
\begin{equation}
X(t) = \mathsf{j}_t(X)=V\left( t\right) ^{\ast } X
  V\left( t\right)
\label{X-qsde}
\end{equation}
and  satisfies
\begin{eqnarray}
 dX(t) =  (\mathcal{L}_{\mathbf{L}(t) } (X(t)) -i [ X(t), H(t) ])dt
\nonumber \\
  +  d\mathbf{A}^\dagger(t) \mathbf{S}^\dagger(t) [ X(t), \mathbf{L}(t)]
+ [\mathbf{L}^\dagger (t),X(t)] \mathbf{S}(t) d\mathbf{A}(t)
\notag \\
  + \mathrm{tr}[ (\mathbf{S}^\dagger (t) X(t) \mathbf{S}(t) - X(t) ) d\mathbf{\Lambda}(t)] .
\label{qle-X}
\end{eqnarray}
 In this expression, all operators evolve unitarily according to (\ref{X-qsde}) (e.g.
$\mathbf{L}(t)=\mathsf{j}_t( \mathbf{L})$)  (commutators
of vectors and matrices of operators are defined component-wise), and tr denotes the trace of a matrix.
We also employ the notation
\begin{eqnarray}
\mathcal{L}_{\mathbf{L}}(X) &=&  \frac{1}{2} \mathbf{L}^\dagger[X,\mathbf{L}]
+ \frac{1}{2} [\mathbf{L}^\dagger, X ] \mathbf{L} 
\notag \\
&=&
\sum_{j=1}^n (  \frac{1}{2} L_j^\ast[X,L_j]
+ \frac{1}{2} [L_j^\ast, X ] L_j) ;
\label{lindblad}
\end{eqnarray}
this is called the {\em Lindblad superoperator} in the physics
literature (it is analogous to the transition matrix for a
classical Markov chain, or the generator of a classical diffusion
process). The dynamics is unitary, and hence preserves commutation
relations. The output fields are defined by
\begin{equation}
\tilde{\mathbf{A}}(t)=V^\ast(t)   \mathbf{A}(t)  V(t)  , \ \  \
\tilde{\mathbf{\Lambda}}(t)=V^\ast(t)   \mathbf{\Lambda}(t)  V(t) ,
\label{open-out-1}
\end{equation}
and satisfy the quantum stochastic differential equations
\begin{eqnarray*}
d\tilde{\mathbf{A}}(t) &=& \mathbf{S}(t) d\mathbf{A}(t) + \mathbf{L}(t) dt
\\
d \tilde {\mathbf{\Lambda}}(t) &=&   \mathbf{S}^\sharp(t)
d\mathbf{\Lambda}(t) \mathbf{S}^T(t) +
  \mathbf{S}^\sharp (t) d\mathbf{A}^\sharp(t) \mathbf{L}^T(t) 
  \\
  &&+ \mathbf{L}^\sharp(t) d\mathbf{A}^T(t) \mathbf{S}^T(t)
  + \mathbf{L}^\sharp(t) \mathbf{L}^T(t) dt ,
\end{eqnarray*}
where $\mathbf{L}(t) = \mathsf{j}_t(\mathbf{L})$, etc, as above.
The output processes also have  canonical quantum It\={o}
products.

In the physics literature, it is common practice to describe open
systems using a {\em master equation} (analogous to the Kolmogorov
equation for the density of a classical diffusion process) for a
density operator $\rho$, a convex combination of outer products
$\psi \psi^\ast$ (here $\psi$ is a state vector). Master equations
can easily be obtained from the parameters $\mathbf{G} =
(\mathbf{S}, \mathbf{L}, H)$; indeed, we have
\begin{equation}
\frac{d}{dt} \rho = i[ \rho, H (t)] + \mathcal{L}'_{\mathbf{L}(t)} (\rho),
\label{master}
\end{equation}
where $\mathcal{L}'_{\mathbf{L}} (\rho )=\mathbf{L}^T  \rho \mathbf{L}^\sharp
- \frac{1}{2} \mathbf{L}^\sharp \mathbf{L}^T  \rho
-   \frac{1}{2} \rho \mathbf{L}^\sharp \mathbf{L}^T $ is the adjoint of the Lindbladian:
 $\mathrm{tr}[ \rho(t)\mathcal{L}_{\mathbf{L}} (X) ] = \mathrm{tr}[\mathcal{L}'_{\mathbf{L}} (\rho) \ X ]$. Note that while the master equation does not depend on the scattering matrix $\mathbf{S}$, this matrix plays an important role in describing the architecture of the input channels, as  in subsections \ref{sec:prelim-series} and \ref{sec:eg-direct-mfb}. We also mention that if an observable of one or more output channels is continuously monitored, then a quantum filter (also called a stochastic master equation) for the conditional density operator can be written down in terms of the parameters $\mathbf{G} = (\mathbf{S}, \mathbf{L}, H)$; an example of this is discussed in subsection \ref{sec:eg-realistic}, see \cite{BHJ07}.

 Open systems specified by parameters $\mathbf{G} = (\mathbf{S}, \mathbf{L}, H)$ preserve the canonical nature of the quantum signals.
However, if  the inputs are not canonical, one will need to modify the equations for the unitary, the Heisenberg dynamics, and the outputs, etc,  to accommodate non-canonical correlations; we do not pursue this matter further here, and in this paper we will always use canonical quantum signals.

\section{The Concatenation and Series Products and their Application to Quantum Networks}
\label{sec:series-apps}

This section contains the main results of the paper. The concatenation and series products are defined in subsection \ref{sec:series-dfns}, and applied to a feedback arrangement in  Theorem \ref{thm:series-fb},  the {\em
principle of series connections} (subsection \ref{sec:series-apps-fb}). This is followed in subsection
\ref{sec:series-apps-cascade} with a specialization to cascade
networks, and a consideration in subsection
\ref{sec:series-apps-reduce} of reducible  networks. These results
are applied to a range of examples in section \ref{sec:eg}.

\subsection{Definitions}
\label{sec:series-dfns}

In this subsection we define two products between system parameters. It is assumed that both systems are defined on the same underlying initial Hilbert space, enlarging if necessary by using a tensor product.

\begin{definition}   \label{dfn:concat} (Concatenation product)
Given two systems $\mathbf{G}_1=(\mathbf{S}_1, \mathbf{L}_1, H_1)$ and $\mathbf{G}_2=(\mathbf{S}_2, \mathbf{L}_2, H_2)$, we define their {\em concatenation} to be the system $\mathbf{G}_1  \boxplus \mathbf{G}_2$ by
\begin{equation}
\mathbf{G}_1  \boxplus \mathbf{G}_2 = (\left( \begin{array}{cc}
\mathbf{S}_1 & 0 \\ 0 & \mathbf{S}_2 \end{array}\right),  \left( \begin{array}{c}  \mathbf{L}_1 \\  \mathbf{L}_2\end{array}\right), H_1+H_2) .
\label{box-plus-dfn}
\end{equation}
\end{definition}

The concatenation product is useful for combining distinct systems, or for decomposing a given system into subsystems. It does not describe interconnections via field channels, but does allow for direct  connections via the Hamiltonian parameters.
Systems without field channels are included   by employing blanks;
    set $\left( \_,\_,H\right) \boxplus \left(
\_,\_,H^{\prime }\right) :=\left( \_,\_,H+H^{\prime }\right) $ and more
generally $\left( \_,\_,H\right) \boxplus \left( S^{\prime },\mathbf{L}%
^{\prime },H^{\prime }\right) =\left( S^{\prime },\mathbf{L}^{\prime
},H^{\prime }\right) \boxplus \left( \_,\_,H\right) :=\left( S^{\prime },%
\mathbf{L}^{\prime },H+H^{\prime }\right) $.

\begin{definition}   \label{dfn:reduce} (Reducible system)
We say that a system  $\mathbf{G}=(\mathbf{S}, \mathbf{L}, H)$ is {\em reducible} if it can be expressed as
\begin{equation}
\mathbf{G} = \mathbf{G}_1  \boxplus \mathbf{G}_2
\label{reducible-dfn-1}
\end{equation}
for two systems $\mathbf{G}_1$ and $\mathbf{G}_2$. In particular, the parameters of a reducible system have the form
\begin{equation}
\mathbf{S} =   \left( \begin{array}{cc}
\mathbf{S}_1 & 0 \\ 0 & \mathbf{S}_2 \end{array}\right), \ \ \
\mathbf{L} = \left( \begin{array}{c}  \mathbf{L}_1 \\  \mathbf{L}_2\end{array}\right), \ \  \
H= H_1+H_2.
\label{reducible-dfn-2}
\end{equation}
Such decompositions are not unique. Furthermore, if one or more of the subsystems is reducible, the reduction process may be iterated to obtain a decomposition $\mathbf{G}= \boxplus_j \mathbf{G}_j$.
\end{definition}

\begin{definition}   \label{dfn:series}
(Series product)
Given two systems  $\mathbf{G}_1=(\mathbf{S}_1, \mathbf{L}_1, H_1)$ and $\mathbf{G}_2=(\mathbf{S}_2, \mathbf{L}_2, H_2)$ with  the same number of field channels,   the {\em series product} $\mathbf{G}_2 \triangleleft \mathbf{G}_1$ defined by
\begin{eqnarray}
\mathbf{G}_2 \triangleleft \mathbf{G}_1  &=&
\label{series-dfn}
  (\mathbf{S}_2\mathbf{S}_1, \mathbf{L}_2+\mathbf{S}_2\mathbf{L}_1, 
  \notag \\
  && H_1+H_2+\frac{1}{2i}(\mathbf{L}_2^\dagger \mathbf{S}_2 \mathbf{L}_1 -\mathbf{L}_1^\dagger \mathbf{S}_2^\dagger \mathbf{L}_2 )).
\notag
\end{eqnarray}
\end{definition}

As will be explained in the following subsection, the series
product specifies the parameters for a system formed by feeding
the output channel of the first system into the input channel of
the second. Both of these products are powerful tools for
describing quantum networks.

\begin{remark}
Let $d G_j (t)$ denote the infinitesimal It\={o} generators
corresponding to parameters $\mathbf{G}_j=(\mathbf{S}_j,
\mathbf{L}_j, H_j)$, for $j=1,2$ respectively, as constructed in
(\ref{open-hp-G}). The generator corresponding to $\mathbf{G}_2
\triangleleft \mathbf{G}_1 $ is then
\begin{equation}
d G (t) = dG_1 (t) + dG_2 (t) + dG_2 (t) dG_1 (t) .
\label{series-diff}
\end{equation}
The last term is to be computed using the It\={o} table for second
order products of differentials.
\end{remark}

\subsection{Feedback}
\label{sec:series-apps-fb}

Let us consider a   reducible  system
 $\mathbf{G}=\mathbf{G}_{1}\boxplus
\mathbf{G}_{2}$ (recall Definition \ref{dfn:reduce}), where number
of channels in the factors is the same (i.e. dim $\mathbf{L}_1$ =
dim $\mathbf{L}_2$). The setup is
sketched in Figure \ref{fig:series-apps-fb-1}. We investigate what will happen if we feed one of the outputs, say $%
\tilde{\mathbf{A}}_{1}$ back in as the input $\mathbf{A} _{2}$.
Either of the two diagrams in Figure \ref{fig:series-apps-fb-2}
may serve to describe the resulting feedback system. Note that the
outputs will be different after the feedback connection has been
made.

 \begin{figure}[h]
\begin{center}
%
%
%
%
%
%
%
%
%
\setlength{\unitlength}{.04cm}
\begin{picture}(120,45)
\label{pic8}

\thicklines
\put(45,15){\line(0,1){30}}
\put(45,15){\line(1,0){30}}
\put(75,15){\line(0,1){30}}
\put(45,45){\line(1,0){30}}

\thinlines
\put(15,35){\vector(1,0){15}}
\put(15,35){\vector(1,0){90}}
\put(60,35){\circle*{4}}

\put(15,25){\vector(1,0){15}}
\put(15,25){\vector(1,0){90}}
\put(60,25){\circle*{4}}

\put(63,37){$1$}
\put(63,18){$2$}

\put(0,25){$\mathbf{A}_2$}  
\put(112,24){$\tilde{\mathbf{A}}_2$}
\put(0,35){$\mathbf{A}_1$}  
\put(112,34){$\tilde{\mathbf{A}}_1$}

\end{picture}
%

\caption{Reducible system $\mathbf{G}_{1}\boxplus
\mathbf{G}_{2}$ with inputs $\mathbf{A}_1, \mathbf{A}_2$ and outputs $\tilde{\mathbf{A}}_1, \tilde{\mathbf{A}}_2$.
  }
\label{fig:series-apps-fb-1}
\end{center}
\end{figure}
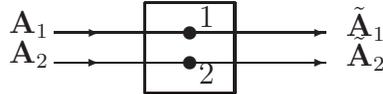

\begin{figure}[h]
\begin{center}
%
%
%
%
%
%
%
%
\setlength{\unitlength}{.04cm}
\begin{picture}(120,40)
\label{pic9a}

\put(63,37){$1$}
\put(63,18){$2$}

\thicklines
\put(45,15){\line(0,1){30}}
\put(45,15){\line(1,0){30}}
\put(75,15){\line(0,1){30}}
\put(45,45){\line(1,0){30}}

\thinlines
\put(15,35){\vector(1,0){15}}
\put(15,35){\line(1,0){70}}
\put(60,35){\circle*{4}}
\put(85,35){\line(0,-1){8}}
\put(85,23){\line(0,-1){23}}
\put(15,0){\line(1,0){70}}
\put(15,0){\line(0,1){25}}

\put(15,25){\vector(1,0){15}}
\put(15,25){\vector(1,0){90}}
\put(60,25){\circle*{4}}

\put(112,24){$\tilde{\mathbf{A}}_2$}
\put(5,35){$\mathbf{A}_1$}
\put(45,5){$\mathbf{A}_2 = \tilde{\mathbf{A}}_1$}

\end{picture}
%
%
%
%
%
%
%
%
%
%
\setlength{\unitlength}{.04cm}
\begin{picture}(120,60)
\label{pic9b}

\thinlines
\put(63,37){$1$}
\put(53,18){$2$}

\thicklines
\put(45,15){\line(0,1){30}}
\put(45,15){\line(1,0){30}}
\put(75,15){\line(0,1){30}}
\put(45,45){\line(1,0){30}}

\thinlines
\put(15,35){\vector(1,0){15}}
\put(15,35){\line(1,0){45}}
\put(60,35){\circle*{4}}
\put(60,25){\line(0,1){10}}

\put(60,25){\vector(1,0){45}}
\put(60,25){\circle*{4}}

\put(112,24){$\tilde{\mathbf{A}}_2$}
\put(5,35){$\mathbf{A}_1$}

\end{picture}
%

\caption{Direct feedback system $\mathbf{G}_{2}\triangleleft
\mathbf{G}_{1}$, with input $\mathbf{A}_1$ and output $\tilde{\mathbf{A}}_2$.
 }
\label{fig:series-apps-fb-2}
\end{center}
\end{figure}
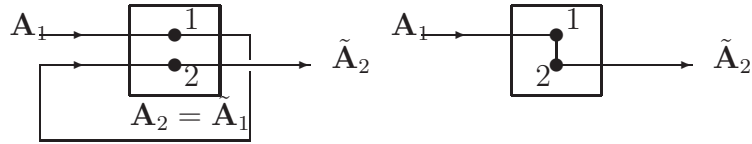

We now state our main result applying the series product to feedback.

\begin{theorem}   \label{thm:series-fb}
\textbf{(Principle of Series Connections) }\textit{The parameters }$\mathbf{G}_{2\leftarrow 1}
$ \textit{for the feedback system obtained from }$\mathbf{G}_{1}\boxplus
\mathbf{G}_{2}$\textit{\ when the output of the  first subsystem is
fed into the input of the
second is given by the series product }$\mathbf{G}_{2\leftarrow 1}=\mathbf{G}%
_{2}\triangleleft \mathbf{G}_{1}$.
\end{theorem}

A proof of this theorem  is given in  Appendix \ref{sec:app-series}.

\subsection{Cascade}
\label{sec:series-apps-cascade}

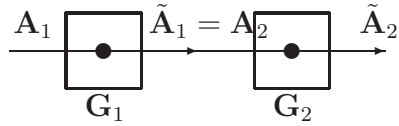
\begin{figure}[h]
\begin{center}
%
%
%
%
%
%
%
%
\setlength{\unitlength}{.05cm}
\begin{picture}(110,40)
\label{pic10}

\newsavebox{\cascadeB}
\savebox{\cascadeB}(10,0){
\thicklines
\put(15,15){\line(0,1){20}}
\put(15,15){\line(1,0){20}}
\put(15,35){\line(1,0){20}}
\put(35,15){\line(0,1){20}}
\thinlines
\put(0,25){\line(1,0){15}}
\put(0,25){\line(1,0){25}}
\put(25,25){\vector(1,0){25}}
\put(25,25){\circle*{4}}}

\put(0,15){\usebox{\cascadeB}}
\put(50,15){\usebox{\cascadeB}}

\put(25,5){$\mathbf{G}_1$}
\put(75,5){$\mathbf{G}_2$}

\put(6,27){$\mathbf{A}_1$}
\put(42,27){${\tilde{\mathbf{A}}}_1=\mathbf{A}_2$}

\put(98,27){${\tilde {\mathbf{A}}}_2$}

\end{picture}
%

\caption{Cascade of independent  quantum components, $\mathbf{G}_2 \triangleleft \mathbf{G}_1$.}
\label{fig:series-apps-cascade-1}
\end{center}
\end{figure}

In our treatment of series connections, we nowhere assumed that the
matrix entries commuted, and this of course facilitated feedback.
However, the principle of series connections also applies to the
special case where the subsystems commute, as in a {\em cascade}
of independent systems, as shown in Figure
\ref{fig:series-apps-cascade-1}. \footnote{Indeed, the reason we
use the term \lq\lq{series}\rq\rq \ is to indicate that it applies
more generally than to cascades of independent components.}

To formulate the cascade arrangement, we first consider the
concatenation of the two systems $\mathbf{G}_1 \boxplus
\mathbf{G}_2$.
The system $\mathbf{G}=\mathbf{G}_{1}\boxplus \mathbf{G}_{2}$  is reducible with components $%
\mathbf{G}_{j}$.

The notion of cascaded quantum systems goes back to Carmichael \cite{HJC93}, who used a quantum trajectory analysis, and Gardiner \cite{CWG93}
who used (scalar) quantum noise models of the form $\mathbf{G}%
_{j} = \left( 1,L_{j},H_{j}\right) $ (no scattering).
 As a special
case of the series principle, we see that the cascaded generator for this
type of setup is $\mathbf{G}_{\text{cascade}} = \mathbf{G}_2 \triangleleft \mathbf{G}_1 = \left(
1,L_{1}+L_{2},H_{1}+H_{2}+\mathrm{Im}\left\{ L_{2}^\ast L_{1}\right\} \right) $%
. This is entirely in agreement with Gardiner's analysis, cf. \cite[Chapter 12]{GZ00} with $L_{j}=\sqrt{\gamma _{j}}c_{j}$ where we have $%
L_{2\leftarrow 1}=\sqrt{\gamma _{1}}c_{1}+\sqrt{\gamma _{2}}c$ and $%
H_{2\leftarrow 1}=H_{1}+H_{2}+\frac{1}{2i }\sqrt{\gamma _{1}\gamma
_{2}}\left( c_{2}^{\ast }c_{1}-c_{1}^{\ast }c_{2}\right) $.

We now consider cascade arrangements and ask what happens if we
try to swap the order of the components. Since the series product
is not in general commutative, we cannot expect to be able to swap
the order without, say, modifying one of the components. We now
make this precise as follows.

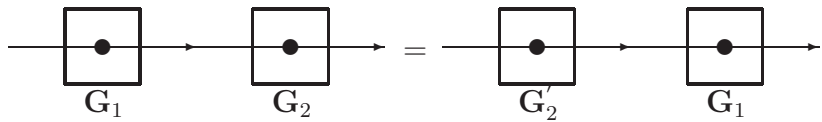
\begin{figure}[h]
\begin{center}
%
%
%
%
%
%
\setlength{\unitlength}{.05cm}
\begin{picture}(110,40)
\label{pic11a}

\put(110,19){$=$}
\newsavebox{\cascadeE}
\savebox{\cascadeE}(10,0){
\thicklines
\put(15,15){\line(0,1){20}}
\put(15,15){\line(1,0){20}}
\put(15,35){\line(1,0){20}}
\put(35,15){\line(0,1){20}}
\thinlines
\put(0,25){\line(1,0){15}}
\put(0,25){\line(1,0){25}}
\put(25,25){\vector(1,0){25}}
\put(25,25){\circle*{4}}}

\put(0,15){\usebox{\cascadeE}}
\put(50,15){\usebox{\cascadeE}}

\put(25,5){${\mathbf G}_1$}
\put(75,5){${\mathbf G}_2$}

\end{picture}
%
%
%
%
%
%
%
%
\setlength{\unitlength}{.05cm}
\begin{picture}(110,40)
\label{pic11b}

\newsavebox{\cascadeF}
\savebox{\cascadeF}(10,0){
\thicklines
\put(15,15){\line(0,1){20}}
\put(15,15){\line(1,0){20}}
\put(15,35){\line(1,0){20}}
\put(35,15){\line(0,1){20}}
\thinlines
\put(0,25){\line(1,0){15}}
\put(0,25){\line(1,0){25}}
\put(25,25){\vector(1,0){25}}
\put(25,25){\circle*{4}}}

\put(0,15){\usebox{\cascadeF}}
\put(50,15){\usebox{\cascadeF}}

\put(25,5){${\mathbf G}_2^{'}$}
\put(75,5){${\mathbf G}_1$}

\end{picture}
%

\caption{Equivalent Systems.}
\label{sec:series-apps-cascade-2}
\end{center}
\end{figure}

We say that two systems are {\em parametrically equivalent} if their parameters are identical. This implies that,
for the same input, they produce
the same internal dynamics and output. Consider the cascaded systems shown in Figure \ref{sec:series-apps-cascade-2}.

We assume that the initial inputs are canonical in both cases and ask, for
fixed choices of $\mathbf{G}_{1}$ and $\mathbf{G}_{2}$, what we should take
for $\mathbf{G}_{2}^{\prime }$ so that the setups are parametrically equivalent.

\begin{theorem}  \label{thm:exchange}
The two  cascaded systems shown in Figure \ref{sec:series-apps-cascade-2} are parametrically equivalent if and only if
\begin{equation}
\mathbf{G_2} \triangleleft \mathbf{G}_1 = \mathbf{G_1} \triangleleft \mathbf{G}_2' .
\label{exchange-1}
\end{equation}
Furthermore, if $\left( \mathbf{S}_{j},\mathbf{L}_{j},H_{j}\right)
$ are the parameters for $\mathbf{G}_{1}$ and $\mathbf{G}_{2}$
$\left( j=1,2\right) $, then the parameters $\left(
\mathbf{S}_{2}^{\prime },\mathbf{L}_{2}^{\prime },H_{2}^{\prime
}\right) $ of $\mathbf{G}_{2}^{\prime }$ are uniquely determined
by
\begin{eqnarray}
\mathbf{S}_{2}^{\prime } &=&\mathbf{S}_{1}^{\dag }\mathbf{S}_{2}\mathbf{S}_{1},  \nonumber \\
\mathbf{L}_{2}^{\prime } &=&\mathbf{S}_{1}^{\dag }\left( \mathbf{S}_{2}-\mathbf{I}\right) \mathbf{L}_{1}+\mathbf{S}_{1}^{\dag
}\mathbf{L}_{2},  \notag \\
H_{2}^{\prime } &=&H_{2}+\mathrm{Im}\left\{ \mathbf{L}_{2}^{\dag }\left( \mathbf{S}_{2}+\mathbf{I}\right)
\mathbf{L}_{1}-\mathbf{L}_{1}^{\dag }\mathbf{S}_{2}\mathbf{L}_{1}\right\} .
\label{exchange-2}
\end{eqnarray}
\end{theorem}

The proof of this theorem is given in
Appendix \ref{sec:app-swap}.

\begin{remark}     \label{rmk:pass-thru}
A useful special case of this result is moving a scattering matrix from the input to the output of a modified system:
\begin{equation}
(\mathbf{S}, \mathbf{L}, H) =
(\mathbf{I}, \mathbf{L}, H) \triangleleft (\mathbf{S}, 0,0) = (\mathbf{S}, 0,0) \triangleleft (\mathbf{I},\mathbf{S}^\dagger  \mathbf{L}, H) .
\label{pass-thru}
\end{equation}
This is illustrated in subsection \ref{sec:prelim-series}.
$\Box$
\end{remark}

\subsection{Reducible Networks}
\label{sec:series-apps-reduce}

Networks can be formed by combining components with the concatenation and series products. Within this framework, components may interact directly, or indirectly via fields. This framework  is  useful for modeling existing systems, as we have seen above, as well as for designing new systems.

Let $\{ \mathbf{G}_j \}$ be a collection of components, which we may combine together to form an unconnected system $\mathbf{G}=\boxplus_j \mathbf{G}_j$. The components may interact directly via bidirectional exchanges of energy, and this may be specified by
a   direct connection Hamiltonian $K$ of the form
\begin{equation}
K = i \sum_k ( N_k^\ast M_k - M_k^\ast N_k) ,
\label{G-reducible-decomp-direct}
\end{equation}
where $M_k$, $N_k$  are operators defined on the initial Hilbert space for $\mathbf{G}$. The components may also interact via field interconnects, specified by a list of series connections
\begin{equation}
\mathscr{S}=   \{  \mathbf{G}_{j_1} \triangleleft \mathbf{G}_{k_1}, \ldots,   \mathbf{G}_{j_n} \triangleleft \mathbf{G}_{k_n} \}
\label{G-reducible-decomp-field}
\end{equation}
such that (i) the field dimensions of the members
of each pair are the same, and (ii) each input and each output
(relative to the decomposition $\mathbf{G}= \boxplus_j \mathbf{G}_j$) has at
most one connection.

\begin{figure}[h]
\begin{center}

\setlength{\unitlength}{1900sp}
\begingroup\makeatletter\ifx\SetFigFont\undefined%
\gdef\SetFigFont#1#2#3#4#5{%
  \reset@font\fontsize{#1}{#2pt}%
  \fontfamily{#3}\fontseries{#4}\fontshape{#5}%
  \selectfont}%
\fi\endgroup%
\begin{picture}(8744,3700)(-1221,-4939)
\put(5401,-4861){\makebox(0,0)[lb]{\smash{{\SetFigFont{7}{8.4}{\familydefault}{\mddefault}{\updefault}{\color[rgb]{0,0,0}$\mathbf{N}$}%
}}}}
{\color[rgb]{0,0,0}\thicklines
\put(5476,-2461){\circle*{168}}
}%
{\color[rgb]{0,0,0}\put(5476,-3211){\circle*{168}}
}%
{\color[rgb]{0,0,0}\put(5476,-3961){\circle*{168}}
}%
{\color[rgb]{0,0,0}\put(751,-3961){\circle*{168}}
}%
{\color[rgb]{0,0,0}\put(751,-3211){\circle*{168}}
}%
{\color[rgb]{0,0,0}\put(751,-2461){\circle*{168}}
}%
{\color[rgb]{0,0,0}\put(751,-1711){\circle*{168}}
}%
{\color[rgb]{0,0,0}\put(3601,-1711){\vector( 1, 0){3900}}
}%
{\color[rgb]{0,0,0}\put(4801,-4411){\framebox(1500,3150){}}
}%
{\color[rgb]{0,0,0}\put(4276,-2461){\vector( 1, 0){525}}
}%
{\color[rgb]{0,0,0}\put(6676,-3211){\vector(-1, 0){375}}
}%
{\color[rgb]{0,0,0}\put(4276,-3961){\vector( 1, 0){525}}
}%
{\color[rgb]{0,0,0}\put(  1,-4411){\framebox(1500,3150){}}
}%
{\color[rgb]{0,0,0}\put(-1199,-3961){\vector( 1, 0){3900}}
}%
{\color[rgb]{0,0,0}\put(-1199,-2461){\vector( 1, 0){3900}}
}%
{\color[rgb]{0,0,0}\put(-1199,-1711){\vector( 1, 0){3900}}
}%
{\color[rgb]{0,0,0}\put(3601,-2461){\line( 1, 0){3900}}
\put(7501,-2461){\line( 0,-1){750}}
\put(7501,-3211){\line(-1, 0){3900}}
\put(3601,-3211){\line( 0,-1){750}}
\put(3601,-3961){\vector( 1, 0){3900}}
}%
{\color[rgb]{0,0,0}\put(-524,-3961){\vector( 1, 0){525}}
}%
{\color[rgb]{0,0,0}\put(-524,-2461){\vector( 1, 0){525}}
}%
{\color[rgb]{0,0,0}\put(-524,-1711){\vector( 1, 0){525}}
}%
{\color[rgb]{0,0,0}\put(2701,-3211){\vector(-1, 0){3900}}
}%
{\color[rgb]{0,0,0}\put(2026,-3211){\vector(-1, 0){525}}
}%
\put(5776,-1936){\makebox(0,0)[lb]{\smash{{\SetFigFont{7}{8.4}{\familydefault}{\mddefault}{\updefault}{\color[rgb]{0,0,0}1}%
}}}}
\put(5701,-2686){\makebox(0,0)[lb]{\smash{{\SetFigFont{7}{8.4}{\familydefault}{\mddefault}{\updefault}{\color[rgb]{0,0,0}2}%
}}}}
\put(5701,-3436){\makebox(0,0)[lb]{\smash{{\SetFigFont{7}{8.4}{\familydefault}{\mddefault}{\updefault}{\color[rgb]{0,0,0}3}%
}}}}
\put(5701,-4186){\makebox(0,0)[lb]{\smash{{\SetFigFont{7}{8.4}{\familydefault}{\mddefault}{\updefault}{\color[rgb]{0,0,0}4}%
}}}}
\put(976,-1936){\makebox(0,0)[lb]{\smash{{\SetFigFont{7}{8.4}{\familydefault}{\mddefault}{\updefault}{\color[rgb]{0,0,0}1}%
}}}}
\put(976,-2761){\makebox(0,0)[lb]{\smash{{\SetFigFont{7}{8.4}{\familydefault}{\mddefault}{\updefault}{\color[rgb]{0,0,0}2}%
}}}}
\put(976,-3511){\makebox(0,0)[lb]{\smash{{\SetFigFont{7}{8.4}{\familydefault}{\mddefault}{\updefault}{\color[rgb]{0,0,0}3}%
}}}}
\put(976,-4186){\makebox(0,0)[lb]{\smash{{\SetFigFont{7}{8.4}{\familydefault}{\mddefault}{\updefault}{\color[rgb]{0,0,0}4}%
}}}}
\put(676,-4861){\makebox(0,0)[lb]{\smash{{\SetFigFont{7}{8.4}{\familydefault}{\mddefault}{\updefault}{\color[rgb]{0,0,0}$\mathbf{G}$}%
}}}}
{\color[rgb]{0,0,0}\put(5476,-1711){\circle*{168}}
}%
\end{picture}%

\caption{A reducible network $\mathbf{N}= \mathbf{G}_1 \boxplus (
\mathbf{G}_4 \triangleleft \mathbf{G}_3 \triangleleft \mathbf{G}_2
)$ formed from the collection $\mathbf{G}  =\mathbf{G}_1\boxplus \mathbf{G}_2 \boxplus \mathbf{G}_3 \boxplus \mathbf{G}_4$ of components with connections
specified by the list of series connections $\mathscr{S}=\{ \mathbf{G}_3
\triangleleft \mathbf{G}_2 , \mathbf{G}_4 \triangleleft
\mathbf{G}_3  \}$.} \label{fig:net-1}
\end{center}
\end{figure}
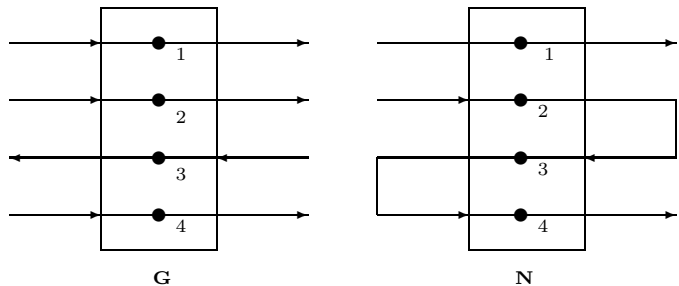

A {\em reducible network} $\mathbf{N}$ is the system formed from $\mathbf{G}$ by implementing the connections (\ref{G-reducible-decomp-direct}) and (\ref{G-reducible-decomp-field}). The parameters of the network $\mathbf{N}$ may be obtained as follows.
A {\em series chain} is a system of the form
$$
\mathbf{C} = \mathbf{G}_{j_l} \triangleleft \mathbf{G}_{k_l}  \triangleleft  \cdots \triangleleft   \mathbf{G}_{j_m} \triangleleft \mathbf{G}_{k_m} .
$$
Let $\mathscr{C}$ denote the set of maximal-length chains drawn from the list of series connections (\ref{G-reducible-decomp-field}), and let $\mathscr{U}$ denote the set of components not involved in any series connection. Then the reducible network is given by
\begin{equation}
\mathbf{N} =   \left(   \boxplus_{\mathbf{G}_k \in \mathscr{U}}  \mathbf{G}_k  \right)  \boxplus  \left(   \boxplus_{\mathbf{C}_j \in \mathscr{C}}  \mathbf{C}_j  \right)   \boxplus (1,0,K) .
\label{N-def}
\end{equation}
 An example of a reducible network is shown in Figure \ref{fig:net-1}.

\begin{remark}    \label{rmk:non-reducible-networks}
The examples considered in section \ref{sec:eg} below are all important examples of reducible networks that have appeared in the literature. However,  we mention that there are important examples of quantum networks
that are not reducible. An example of a non-reducible network was
considered by Yanagisawa and Kimura, \cite[Fig. 4]{YK03a}, which
consists of two systems in a feedback arrangement formed by a beam
splitter, as occurs if in Figure \ref{fig:cavity-bs-1} we connect
the output $\tilde{B}_1$ to the input $A_2$ (i.e. setting
$A_2=\tilde{B}_1$). The feedback loop formed in this way is
\lq\lq{algebraic}\rq\rq, and the resulting in-loop field  is not a
free field in general.
A general theory of quantum feedback networks, both reducible and non-reducible, is given in \cite{GJ08a}.
$\Box$
\end{remark}

\section{Examples}
\label{sec:eg}

In this section we look at a number of examples from the literature which can be represented by reducible networks.

\subsection{All-Optical Feedback}
\label{sec:eg-alloptical}

\begin{figure}[h]
\begin{center}

 \begin{picture}(0,0)%
\includegraphics{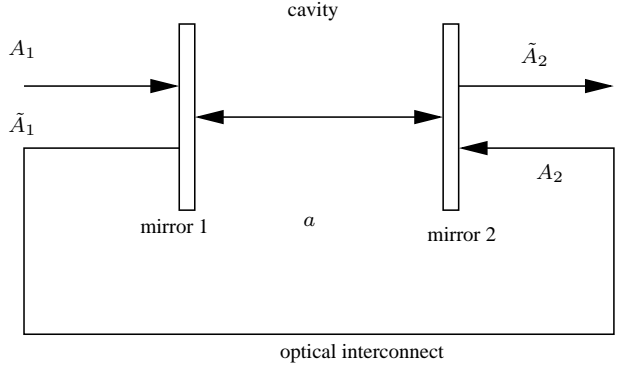}%
\end{picture}%
\setlength{\unitlength}{2565sp}%
\begingroup\makeatletter\ifx\SetFigFont\undefined%
\gdef\SetFigFont#1#2#3#4#5{%
  \reset@font\fontsize{#1}{#2pt}%
  \fontfamily{#3}\fontseries{#4}\fontshape{#5}%
  \selectfont}%
\fi\endgroup%
\begin{picture}(5887,3544)(1936,-5759)
\put(4801,-4411){\makebox(0,0)[lb]{\smash{{\SetFigFont{8}{9.6}{\familydefault}{\mddefault}{\updefault}{\color[rgb]{0,0,0}$a$}%
}}}}
\put(1951,-2761){\makebox(0,0)[lb]{\smash{{\SetFigFont{8}{9.6}{\familydefault}{\mddefault}{\updefault}{\color[rgb]{0,0,0}$A_1$}%
}}}}
\put(1951,-3511){\makebox(0,0)[lb]{\smash{{\SetFigFont{8}{9.6}{\familydefault}{\mddefault}{\updefault}{\color[rgb]{0,0,0}$\tilde{A}_1$}%
}}}}
\put(7051,-3961){\makebox(0,0)[lb]{\smash{{\SetFigFont{8}{9.6}{\familydefault}{\mddefault}{\updefault}{\color[rgb]{0,0,0}$A_2$}%
}}}}
\put(6901,-2836){\makebox(0,0)[lb]{\smash{{\SetFigFont{8}{9.6}{\familydefault}{\mddefault}{\updefault}{\color[rgb]{0,0,0}$\tilde{A}_2$}%
}}}}
\end{picture}%

\caption{All-optical feedback for a cavity. The feedback path is a light beam
from mirror 1 to mirror 2,  both of which are partially transmitting). There is a phase
shift $\theta$ along the feedback path.}
\label{fig:eg-all-optical-1}
\end{center}
\end{figure}

We consider a simple situation first introduced by Wiseman and
Milburn as an example of all-optical feedback, \cite[section II.B.
A]{WM94b}. Referring to Figure \ref{fig:eg-all-optical-1}, vacuum
light field $A_1$ is reflected off mirror 1 to yield an output
beam $\tilde{A}_1$ which results from interaction with the
internal cavity mode $a$. This beam is reflected onto mirror 2, as
shown, where it constitutes the input $A_2$.  It is assume that
both mirrors have the same transmittivity, so that we can model
the coupling operators for the two field channels as
$L_1=L_2=\sqrt{\gamma} \, a$, where $\gamma$ is the damping rate.
We may also assume that the light picks up a phase $S=e^{i \theta
}$ when reflected by the cavity mirror.

Before feedback, the cavity is described by
$$
\mathbf{G}  = (\mathbf{I}, \left( \begin{array}{c} L_1 \\ L_2 \end{array} \right), 0) = (1, L_1, 0) \boxplus (1, L_2, 0) .
$$
The phase shift between the mirrors is described by the system $(S,0,0)$.

\begin{figure}[h]
\begin{center}
%
%
%
%
%
%
%
%
%
%
\setlength{\unitlength}{.04cm}
\begin{picture}(120,60)
\label{pic19a}

\put(63,37){$L_1$}
\put(63,18){$L_2$}

\thicklines
\put(45,15){\line(0,1){30}}
\put(45,15){\line(1,0){30}}
\put(75,15){\line(0,1){30}}
\put(45,45){\line(1,0){30}}

\thinlines
\put(15,35){\vector(1,0){15}}
\put(15,35){\line(1,0){70}}
\put(60,35){\circle*{4}}
\put(85,35){\line(0,-1){8}}
\put(85,23){\line(0,-1){23}}
\put(15,0){\line(1,0){70}}
\put(15,0){\line(0,1){25}}

\put(15,25){\vector(1,0){15}}
\put(15,25){\vector(1,0){90}}
\put(60,25){\circle*{4}}

\put(112,24){$\tilde{A}_2$}
\put(5,40){$A_1$}

\put(60,0){\circle*{4}}

\put(55,-5){\line(1,0){10}}
\put(55,-5){\line(0,1){10}}
\put(65,-5){\line(0,1){10}}
\put(55,5){\line(1,0){10}}
\put(69,-9){$(S,0,0)$}

\end{picture}
%
\qquad
%
%
%
%
%
%
%
%
%
%
\setlength{\unitlength}{.04cm}
\begin{picture}(120,60)
\label{pic19b}

\put(63,37){$L_1'$}
\put(63,18){$L_2'$}

\thicklines
\put(45,15){\line(0,1){30}}
\put(45,15){\line(1,0){30}}
\put(75,15){\line(0,1){30}}
\put(45,45){\line(1,0){30}}

\thinlines
\put(15,35){\vector(1,0){15}}
\put(15,35){\line(1,0){70}}
\put(60,35){\circle*{4}}
\put(85,35){\line(0,-1){8}}
\put(85,23){\line(0,-1){23}}
\put(15,0){\line(1,0){70}}
\put(15,0){\line(0,1){25}}

\put(15,25){\vector(1,0){15}}
\put(15,25){\vector(1,0){110}}
\put(60,25){\circle*{4}}

\put(112,34){$\tilde{A}_2$}
\put(5,40){$A_1$}

\put(100,25){\circle*{4}}

\put(95,20){\line(1,0){10}}
\put(95,20){\line(0,1){10}}
\put(105,20){\line(0,1){10}}
\put(95,30){\line(1,0){10}}
\put(90,10){$(S,0,0)$}

\end{picture}
%

\bigskip

\caption{Representations of the all-optical feedback scheme of Figure \ref{fig:eg-all-optical-1} as  reducible networks.}
\label{fig:eg-all-optical-2}
\end{center}
\end{figure}
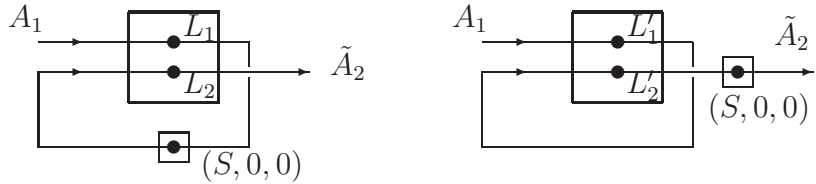

Two equivalent reducible network representations are shown in Figure \ref{fig:eg-all-optical-2}. From the left diagram in
 Figure \ref{fig:eg-all-optical-2}, we see that the closed loop system is described by
 \begin{eqnarray*}
 \mathbf{G}_{cl} &=& (1, L_2, 0) \triangleleft (S,0,0) \triangleleft (1,L_1,0)
 \\
 &=& (S, SL_1+L_2, \frac{1}{2i} ( L_2^\ast S L_1-L_1^\ast S^\ast L_2) ) .
 \end{eqnarray*}
 Here we have twice applied the formulas (\ref{series-dfn}) given in Definition \ref{dfn:series}.

 Alternatively, we may use our theory of equivalent components (Theorem \ref{thm:exchange}) to move the phase
change $\left( S,0,0\right) $ to the very end, as shown in the  right diagram in
 Figure \ref{fig:eg-all-optical-2}.  Then
 \begin{eqnarray*}
 \mathbf{G}_{cl} &=&  (S,0,0) \triangleleft (1, S^\ast L_2, 0) \triangleleft (1,L_1,0)
 \\
 &=& (S, SL_1+L_2, \frac{1}{2i} ( L_2^\ast S L_1-L_1^\ast S^\ast L_2) ) ,
 \end{eqnarray*}
as before.
Either way, the closed loop feedback system is
described by $\mathbf{G}_{cl}=\left( S_{\text{cl}},L_{\text{cl}},H_{\text{cl}}\right) $ where
\begin{eqnarray*}
S_{\text{cl}} &=&S\equiv e^{i \theta }, \\
L_{\text{cl}} &=&S L_{1}+L_{2}  \equiv \left( 1+e^{i\theta }\right) \sqrt{\gamma }a, \\
H_{\text{cl}} &=&\mathrm{Im}\left\{ L_2^{\ast }SL_1\right\} \equiv \gamma \sin
\theta \,a^{\dag }a.
\end{eqnarray*}
From this we obtain the Heisenberg dynamical equation for the cavity mode
\begin{eqnarray*}
d a &=&- \left[ a,\left( 1+e^{i\theta }\right) \sqrt{\gamma
}a^{\dag }\right] dA_{1}
\notag \\
&& - \frac{\gamma }{2}\left( 1+e^{i \theta
}\right) \left( 1+e^{-i\theta }\right)  a dt -i\gamma \sin \theta \, a  dt \\
&\equiv &-\left( 1+e^{i \theta }\right) \left( \sqrt{\gamma }dA_{1}+\gamma a dt\right) ,
\end{eqnarray*}
and the input/output relation, in agreement with \cite[eq.
(2.29)]{WM94b},
\begin{equation*}
d \tilde{A}_2=e^{i\theta }dA_{1}+\left( 1+e^{i\theta }\right) \sqrt{\gamma }a dt.
\end{equation*}

\subsection{Direct Measurement  Feedback}
\label{sec:eg-direct-mfb}

In the paper \cite{HW94a}, Wiseman considers two types of
measurement feedback, one involving photon counting, and another
based on  quadrature measurement using homodyne detection (which
is a diffusive limit of photon counts). In both cases proportional
feedback involving an electrical current was used. We describe
these feedback situations in the following subsections using our
network theory.

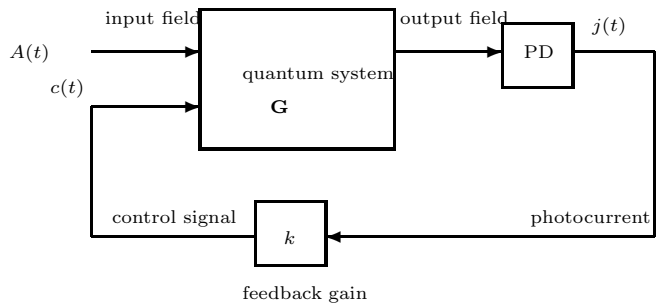
\begin{figure}[h]
\begin{center}
\setlength{\unitlength}{2268sp}
\begingroup\makeatletter\ifx\SetFigFont\undefined%
\gdef\SetFigFont#1#2#3#4#5{%
  \reset@font\fontsize{#1}{#2pt}%
  \fontfamily{#3}\fontseries{#4}\fontshape{#5}%
  \selectfont}%
\fi\endgroup%
\begin{picture}(7072,3226)(-4349,-5365)
\put(-1799,-5311){\makebox(0,0)[lb]{\smash{{\SetFigFont{7}{8.4}{\familydefault}{\mddefault}{\updefault}{\color[rgb]{0,0,0}feedback gain}%
}}}}
\thicklines
{\color[rgb]{0,0,0}\put(1051,-2986){\framebox(750,675){}}
}%
{\color[rgb]{0,0,0}\put(-1649,-4936){\framebox(750,675){}}
}%
{\color[rgb]{0,0,0}\put(1801,-2611){\line( 1, 0){900}}
\put(2701,-2611){\line( 0,-1){2025}}
\put(2701,-4636){\vector(-1, 0){3600}}
}%
{\color[rgb]{0,0,0}\put(-1649,-4636){\line(-1, 0){1800}}
\put(-3449,-4636){\line( 0, 1){1425}}
}%
{\color[rgb]{0,0,0}\put(-3449,-2611){\vector( 1, 0){1200}}
}%
{\color[rgb]{0,0,0}\put(-149,-2611){\vector( 1, 0){1200}}
}%
{\color[rgb]{0,0,0}\put(-3449,-3211){\vector( 1, 0){1200}}
}%
\put(1276,-2686){\makebox(0,0)[lb]{\smash{{\SetFigFont{7}{8.4}{\familydefault}{\mddefault}{\updefault}{\color[rgb]{0,0,0}PD}%
}}}}
\put(2026,-2386){\makebox(0,0)[lb]{\smash{{\SetFigFont{7}{8.4}{\familydefault}{\mddefault}{\updefault}{\color[rgb]{0,0,0}$j(t)$}%
}}}}
\put(-3224,-4486){\makebox(0,0)[lb]{\smash{{\SetFigFont{7}{8.4}{\familydefault}{\mddefault}{\updefault}{\color[rgb]{0,0,0}control signal}%
}}}}
\put(1351,-4486){\makebox(0,0)[lb]{\smash{{\SetFigFont{7}{8.4}{\familydefault}{\mddefault}{\updefault}{\color[rgb]{0,0,0}photocurrent}%
}}}}
\put(-3299,-2311){\makebox(0,0)[lb]{\smash{{\SetFigFont{7}{8.4}{\familydefault}{\mddefault}{\updefault}{\color[rgb]{0,0,0}input field}%
}}}}
\put(-74,-2311){\makebox(0,0)[lb]{\smash{{\SetFigFont{7}{8.4}{\familydefault}{\mddefault}{\updefault}{\color[rgb]{0,0,0}output field}%
}}}}
\put(-1349,-4711){\makebox(0,0)[lb]{\smash{{\SetFigFont{7}{8.4}{\familydefault}{\mddefault}{\updefault}{\color[rgb]{0,0,0}$k$}%
}}}}
\put(-1799,-2911){\makebox(0,0)[lb]{\smash{{\SetFigFont{7}{8.4}{\familydefault}{\mddefault}{\updefault}{\color[rgb]{0,0,0}quantum system}%
}}}}
\put(-4349,-2686){\makebox(0,0)[lb]{\smash{{\SetFigFont{7}{8.4}{\familydefault}{\mddefault}{\updefault}{\color[rgb]{0,0,0}$A(t)$}%
}}}}
\put(-3899,-3061){\makebox(0,0)[lb]{\smash{{\SetFigFont{7}{8.4}{\familydefault}{\mddefault}{\updefault}{\color[rgb]{0,0,0}$c(t)$}%
}}}}
\put(-1499,-3286){\makebox(0,0)[lb]{\smash{{\SetFigFont{7}{8.4}{\familydefault}{\mddefault}{\updefault}{\color[rgb]{0,0,0}$\mathbf{G}$}%
}}}}
{\color[rgb]{0,0,0}\put(-2249,-3661){\framebox(2100,1500){}}
}%
\end{picture}%

\caption{Direct feedback of photocurrent obtained by photon counting using a photodetector (PD).}
\label{fig:direct-fb-count-1}
\end{center}
\end{figure}

Consider the measurement feedback arrangement
shown in Figure \ref{fig:direct-fb-count-1}, which shows a vacuum
input field $A$, a control signal $c$, a photodetector PD, and a
proportional feedback gain $k$.

Before feedback, the quantum system is described by
\begin{equation}
\mathbf{G}= (1, L, H_0 + Fc) ,
\label{photocount-0}
\end{equation}
where $H_0$ and $F$ are self-adjoint, and $c$ represent a classical control variable.
The  photocurrent $j(t)$ resulting from ideal photodetection of  the output field is given by
\begin{equation}
`` j(t)dt "
 =  d\Lambda + L dA^\dag + L^\dag dA + L^\dag L dt ,
\label{photocount-1}
\end{equation}
where, mathematically,  the photocurrent $j(t)$ is the formal
derivative of a field observable (a self-adjoint commutative jump
stochastic process) $\tilde{\Lambda}(t)$ (the output gauge
process) whose It\={o} differential is given by the RHS of
(\ref{photocount-1}). The feedback is given by
\begin{equation}
c(t) = k j(t),
\label{photocount-2}
\end{equation}
where $k$ is a (real, scalar) proportional gain. The feedback gain
can be absorbed into $F$, and so we assume $k=1$ in what follows.

An alternative is to again consider the quantum system
$\mathbf{G}$ given by (\ref{photocount-0}), but replace the
photodetector PD in Figure \ref{fig:direct-fb-count-1} with a
homodyne detector HD.\footnote{An ideal homodyne detector HD takes
an input field $A$ and produces a quadrature, say $A+A^\ast$ (real quadrature), thus
effecting a measurement. This is achieved routinely to good
accuracy in optics laboratories, \cite[Chapter 8]{GZ00}.}
 The homodyne detector then produces a photocurrent $j(t)$ given by
$$
`` {j(t)dt} " = dJ(t) = (L(t)+L^\sharp (t))dt + dA(t)+dA^\sharp(t)
.
$$
The feedback is given by (\ref{photocount-2}) as above, with
feedback gain   absorbed into $F$, as above. The measurement
result $J(t)$ is a field observable (here a  self-adjoint commutative diffusive process).

In order to describe these types of direct measurement feedback within our framework, we view the setup before feedback as being described by
$$
\mathbf{G} = (1, L, H_0)  \boxplus (S_{fb},L_{fb}, H_{fb}) \equiv
\mathbf{G}_0 \boxplus \mathbf{G}_{fb} .
$$
Here, $\mathbf{G}_0$ describes the internal energy of the system and its coupling to the input field $A$. The second term, $\mathbf{G}_{fb}$, describes the way in which the classical input signal is determined from a second  quantum input field (which will be replaced by the output $\tilde A$ when the feedback loop is closed).
The idea is that by appropriate choice of the coupling operator $L_{fb}$, the relevant observable of the field can be selected. In this way, the photodection and homodyne detection measurements are  accommodated. The singular nature of the feedback signal (which contains white noise in the homodyne case) means that care must be taken to describe it correctly.
The correct form of the parameters is given by the Holevo parameterization (Appendix \ref{sec:TOE}, equation (\ref{hp-holevo})) rather than the expression arising from the implicit-explicit formalism of \cite{HW94a}, since the later does not capture correctly gauge couplings, see Appendix \ref{sec:TOE}. We shall interpret the feedback interaction as being due to a
Holevo generator $K_{fb} (t) =H_{00} t + H_{01} A(t) + H_{10} A^\ast(t)  + H_{11}\Lambda(t)$, see Appendix \ref{sec:TOE}, equation (\ref{H-parameters}).
The closed loop system  after
feedback  is given by the series connection
$
\mathbf{G}_{cl} =   \mathbf{G}_{fb} \triangleleft \mathbf{G}_0 =
\left( S_{fb}, L_{fb} + S_{fb}L, H_0+ H_{fb} + \mathrm{Im} \left(
L_{fb}^\ast S_{fb} L_0 \right) \right)  .
$

\subsubsection{Photon Counting}
\label{sec:eg-direct-mfb-count}

\begin{figure}[h]
\begin{center}

\setlength{\unitlength}{2368sp}%
\begingroup\makeatletter\ifx\SetFigFont\undefined%
\gdef\SetFigFont#1#2#3#4#5{%
  \reset@font\fontsize{#1}{#2pt}%
  \fontfamily{#3}\fontseries{#4}\fontshape{#5}%
  \selectfont}%
\fi\endgroup%
\begin{picture}(5069,2594)(2754,-7133)
\put(5251,-6136){\makebox(0,0)[lb]{\smash{{\SetFigFont{7}{8.4}{\familydefault}{\mddefault}{\updefault}{\color[rgb]{0,0,0}$S$}%
}}}}
\thicklines
{\color[rgb]{0,0,0}\put(4801,-6361){\framebox(1200,1800){}}
}%
{\color[rgb]{0,0,0}\put(4276,-5086){\vector( 1, 0){2775}}
}%
{\color[rgb]{0,0,0}\put(3751,-5086){\vector( 1, 0){600}}
}%
{\color[rgb]{0,0,0}\put(3751,-5836){\vector( 1, 0){600}}
}%
{\color[rgb]{0,0,0}\put(6976,-5086){\line( 1, 0){825}}
\put(7801,-5086){\line( 0,-1){2025}}
\put(7801,-7111){\line(-1, 0){5025}}
\put(2776,-7111){\line( 0, 1){1275}}
\put(2776,-5836){\line( 1, 0){1125}}
}%
{\color[rgb]{0,0,0}\put(4276,-5836){\vector( 1, 0){2775}}
}%
\thinlines
{\color[rgb]{0,0,0}\put(5326,-5911){\framebox(225,225){}}
}%
\put(3301,-4936){\makebox(0,0)[lb]{\smash{{\SetFigFont{7}{8.4}{\familydefault}{\mddefault}{\updefault}{\color[rgb]{0,0,0}$A$}%
}}}}
\put(3301,-5686){\makebox(0,0)[lb]{\smash{{\SetFigFont{7}{8.4}{\familydefault}{\mddefault}{\updefault}{\color[rgb]{0,0,0}$C$}%
}}}}
\put(6751,-5686){\makebox(0,0)[lb]{\smash{{\SetFigFont{7}{8.4}{\familydefault}{\mddefault}{\updefault}{\color[rgb]{0,0,0}$\tilde{C}$}%
}}}}
\put(6751,-4861){\makebox(0,0)[lb]{\smash{{\SetFigFont{7}{8.4}{\familydefault}{\mddefault}{\updefault}{\color[rgb]{0,0,0}$\tilde{A}$}%
}}}}
\put(5251,-4861){\makebox(0,0)[lb]{\smash{{\SetFigFont{7}{8.4}{\familydefault}{\mddefault}{\updefault}{\color[rgb]{0,0,0}$L$}%
}}}}
{\color[rgb]{0,0,0}\thicklines
\put(5401,-5086){\circle*{150}}
}%
\end{picture}%

\caption{Representation of the direct photocount feedback
scheme of Figure \ref{fig:direct-fb-count-1} as a reducible network.}
\label{fig:direct-fb-count-1a}
\end{center}
\end{figure}
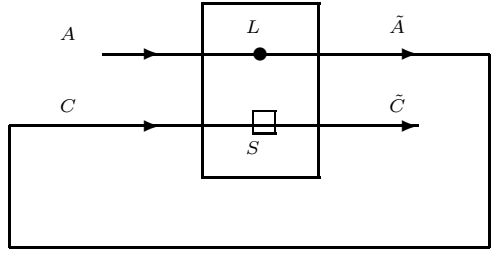

Here we take $K_{fb} (t) = F \Lambda (t)$, so that $S_{fb} =
e^{-iF}$, see Appendix \ref{sec:TOE}, equation (\ref{hp-holevo}). Note that this coupling picks out the required photon number observable  of the field. We then have $ \mathbf{G}_{fb} =
(e^{-iF},0,0)$ and so
$$
\mathbf{G}_{cl} = (e^{-iF},e^{-iF}L,H_0) .
$$
This is illustrated in Figure \ref{fig:direct-fb-count-1a}. The
resulting Heisenberg equation agrees with the results obtained by
Wiseman, \cite[eq. (3.44)]{HW94a}, which we write in our notation
as
\begin{eqnarray}
dX &=&  ( -i [X,  H_0 ] + \mathcal{L}_{e^{-iF}L}(X))dt +(e^{iF} X
e^{-iF} -X)d\Lambda
\notag \\
&&+ e^{iF}
[X,e^{-iF}L] dA^\ast + [L^\ast e^{iF}, X] e^{-iF} dA   .
\label{sys-field-qle-count}
\end{eqnarray}

({\em Technical aside.} Note that if we set $E(t) = E   \Lambda
(t)$, with $E$ self-adjoint, then the Stratonovich  equation
$dV(t) = -i dE(t) \circ V(t) \equiv -idE(t) V(t)
-\frac{i}{2}dE(t)dV(t)$ is equivalent to $dV(t) = S_{fb} d\Lambda
(t) V(t)$ where $S_{fb} = \frac{1 - \frac{i}{2}E}{1 +
\frac{i}{2}E}$. Therefore the implicit form \cite{HW94a}  is not
the Stratonovich form \cite{JG06}.)

\subsubsection{Quadrature Measurement}
\label{sec:eg-direct-mfb-quad}

\begin{figure}[h]
\begin{center}

\setlength{\unitlength}{2368sp}%
\begingroup\makeatletter\ifx\SetFigFont\undefined%
\gdef\SetFigFont#1#2#3#4#5{%
  \reset@font\fontsize{#1}{#2pt}%
  \fontfamily{#3}\fontseries{#4}\fontshape{#5}%
  \selectfont}%
\fi\endgroup%
\begin{picture}(5069,2594)(2754,-7133)
\put(5326,-6211){\makebox(0,0)[lb]{\smash{{\SetFigFont{7}{8.4}{\familydefault}{\mddefault}{\updefault}{\color[rgb]{0,0,0}$M$}%
}}}}
{\color[rgb]{0,0,0}\thicklines
\put(5401,-5836){\circle*{150}}
}%
{\color[rgb]{0,0,0}\put(4801,-6361){\framebox(1200,1800){}}
}%
{\color[rgb]{0,0,0}\put(4276,-5086){\vector( 1, 0){2775}}
}%
{\color[rgb]{0,0,0}\put(3751,-5086){\vector( 1, 0){600}}
}%
{\color[rgb]{0,0,0}\put(4276,-5836){\vector( 1, 0){2775}}
}%
{\color[rgb]{0,0,0}\put(3751,-5836){\vector( 1, 0){600}}
}%
{\color[rgb]{0,0,0}\put(6976,-5086){\line( 1, 0){825}}
\put(7801,-5086){\line( 0,-1){2025}}
\put(7801,-7111){\line(-1, 0){5025}}
\put(2776,-7111){\line( 0, 1){1275}}
\put(2776,-5836){\line( 1, 0){1125}}
}%
\put(3301,-4936){\makebox(0,0)[lb]{\smash{{\SetFigFont{7}{8.4}{\familydefault}{\mddefault}{\updefault}{\color[rgb]{0,0,0}$A$}%
}}}}
\put(3301,-5686){\makebox(0,0)[lb]{\smash{{\SetFigFont{7}{8.4}{\familydefault}{\mddefault}{\updefault}{\color[rgb]{0,0,0}$C$}%
}}}}
\put(6751,-5686){\makebox(0,0)[lb]{\smash{{\SetFigFont{7}{8.4}{\familydefault}{\mddefault}{\updefault}{\color[rgb]{0,0,0}$\tilde{C}$}%
}}}}
\put(6751,-4861){\makebox(0,0)[lb]{\smash{{\SetFigFont{7}{8.4}{\familydefault}{\mddefault}{\updefault}{\color[rgb]{0,0,0}$\tilde{A}$}%
}}}}
\put(5326,-4861){\makebox(0,0)[lb]{\smash{{\SetFigFont{7}{8.4}{\familydefault}{\mddefault}{\updefault}{\color[rgb]{0,0,0}$L$}%
}}}}
{\color[rgb]{0,0,0}\put(5401,-5086){\circle*{150}}
}%
\end{picture}%

\caption{Representation of the direct homodyne feedback  scheme (Figure \ref{fig:direct-fb-count-1} with HD replacing PD) as a reducible network.}
\label{fig:direct-fb-1a}
\end{center}
\end{figure}
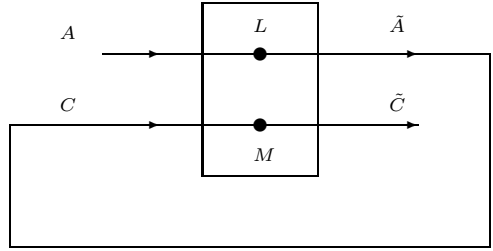

Here we take $K_{fb} (t) = F (A^\ast (t) +A(t))$ in which
case $ \mathbf{G}_{fb} = (1, -iF,0) $, see Appendix \ref{sec:TOE}, equation (\ref{hp-holevo}).
The skew-symmetry of $-iF$ ensures that the coupling selects the desired field quadrature observable.
After
feedback, the closed loop system is
$$
\mathbf{G}_{cl} = (1, L-iF, H_0 + \frac{1}{2} (FL+L^\ast F ))
$$
using (\ref{series-dfn}). This is illustrated in  Figure \ref{fig:direct-fb-1a}.
The resulting Heisenberg equation then agrees with \cite[eq. (4.21)]{HW94a},
which we write as
\begin{eqnarray}
dX &=&  ( -i [X, H_0 + \frac{1}{2}( FL + L^\ast F ) ] +
\mathcal{L}_{L-iF}(X))dt 
\notag \\
&&+ [X,(L-iF)] dA^\ast + [(L-iF)^\ast,
X])dA . 
\label{sys-field-qle-2}
\end{eqnarray}

({\em Technical aside.}  Note that for diffusions (that is, no gauge terms) the Holevo generator and
Stratonovich generator coincide: that is, $dV(t) =
(e^{-idK_{fb}(t)}-1)V(t)$ is the same as $dV(t) = -idK_{fb}
(t)\circ V(t)$, Appendix \ref{sec:TOE}.)

\subsection{Realistic Detection}
\label{sec:eg-realistic}

\begin{figure}[h]
\begin{center}

\begin{picture}(0,0)%
\includegraphics{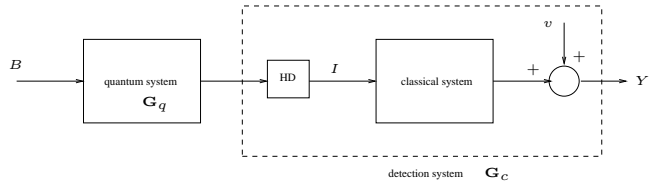}%
\end{picture}%
\setlength{\unitlength}{1381sp}%
\begingroup\makeatletter\ifx\SetFigFont\undefined%
\gdef\SetFigFont#1#2#3#4#5{%
  \reset@font\fontsize{#1}{#2pt}%
  \fontfamily{#3}\fontseries{#4}\fontshape{#5}%
  \selectfont}%
\fi\endgroup%
\begin{picture}(11280,3175)(-3614,-4714)
\put(-1199,-3361){\makebox(0,0)[lb]{\smash{{\SetFigFont{5}{6.0}{\familydefault}{\mddefault}{\updefault}{\color[rgb]{0,0,0}$\mathbf{G}_q$}%
}}}}
\put(6526,-2536){\makebox(0,0)[lb]{\smash{{\SetFigFont{5}{6.0}{\familydefault}{\mddefault}{\updefault}{\color[rgb]{0,0,0}$+$}%
}}}}
\put(5701,-2761){\makebox(0,0)[lb]{\smash{{\SetFigFont{5}{6.0}{\familydefault}{\mddefault}{\updefault}{\color[rgb]{0,0,0}$+$}%
}}}}
\put(7651,-2986){\makebox(0,0)[lb]{\smash{{\SetFigFont{5}{6.0}{\familydefault}{\mddefault}{\updefault}{\color[rgb]{0,0,0}$Y$}%
}}}}
\put(2176,-2761){\makebox(0,0)[lb]{\smash{{\SetFigFont{5}{6.0}{\familydefault}{\mddefault}{\updefault}{\color[rgb]{0,0,0}$I$}%
}}}}
\put(-3599,-2686){\makebox(0,0)[lb]{\smash{{\SetFigFont{5}{6.0}{\familydefault}{\mddefault}{\updefault}{\color[rgb]{0,0,0}$B$}%
}}}}
\put(6001,-1936){\makebox(0,0)[lb]{\smash{{\SetFigFont{5}{6.0}{\familydefault}{\mddefault}{\updefault}{\color[rgb]{0,0,0}$v$}%
}}}}
\put(4951,-4636){\makebox(0,0)[lb]{\smash{{\SetFigFont{5}{6.0}{\familydefault}{\mddefault}{\updefault}{\color[rgb]{0,0,0}$\mathbf{G}_c$}%
}}}}
\end{picture}%

\caption{Model of a realistic detection scheme for a quantum
system, showing ideal homodyne detection followed by a classical
system (e.g. low pass filter) and additive classical noise.}
\label{fig:realistic-1}
\end{center}
\end{figure}

Consider a quantum system $\mathbf{G}_q$ continuously monitored by
observing the real quadrature $\tilde B + \tilde B^\ast$ of an output field $\tilde B$. This measurement
can ideally be carried out by homodyne detection, but due to
finite bandwidth of the electronics and electrical noise, this
measurement could be more accurately modeled by introducing a
classical system (low pass filter)  and additive noise as shown in
Figure \ref{fig:realistic-1},  as analyzed in \cite{WWM02}. Here,
$B$ is a  vacuum field,  $I$ is the output of the ideal homodyne
detector (HD), $v$ is a standard  Wiener process, and $Y$ is the
(integral of) the electric current providing the measurement
information.

We wish to derive a filter to estimate quantum system variables
$X_q$ from the information available in the measurement $Y$.

\begin{figure}[h]
\begin{center}

\setlength{\unitlength}{1900sp}
\begingroup\makeatletter\ifx\SetFigFont\undefined%
\gdef\SetFigFont#1#2#3#4#5{%
  \reset@font\fontsize{#1}{#2pt}%
  \fontfamily{#3}\fontseries{#4}\fontshape{#5}%
  \selectfont}%
\fi\endgroup%
\begin{picture}(8648,2705)(151,-7244)
\put(8476,-5911){\makebox(0,0)[lb]{\smash{{\SetFigFont{7}{8.4}{\familydefault}{\mddefault}{\updefault}{\color[rgb]{0,0,0}$Y$}%
}}}}
{\color[rgb]{0,0,0}\thicklines
\put(5401,-5836){\circle*{150}}
}%
{\color[rgb]{0,0,0}\put(1951,-5086){\circle*{150}}
}%
{\color[rgb]{0,0,0}\put(4801,-6361){\framebox(1200,1800){}}
}%
{\color[rgb]{0,0,0}\put(4276,-5086){\vector( 1, 0){2775}}
}%
{\color[rgb]{0,0,0}\put(3751,-5086){\vector( 1, 0){600}}
}%
{\color[rgb]{0,0,0}\put(4276,-5836){\vector( 1, 0){2775}}
}%
{\color[rgb]{0,0,0}\put(3751,-5836){\vector( 1, 0){600}}
}%
{\color[rgb]{0,0,0}\put(1426,-6361){\framebox(1200,1800){}}
}%
{\color[rgb]{0,0,0}\put(1126,-5086){\line( 1, 0){2775}}
}%
{\color[rgb]{0,0,0}\put(526,-5086){\vector( 1, 0){600}}
}%
{\color[rgb]{0,0,0}\put(7051,-6136){\framebox(600,600){}}
}%
{\color[rgb]{0,0,0}\put(7651,-5836){\vector( 1, 0){750}}
}%
\put(4726,-6811){\makebox(0,0)[lb]{\smash{{\SetFigFont{7}{8.4}{\familydefault}{\mddefault}{\updefault}{\color[rgb]{0,0,0}classical system}%
}}}}
\put(3301,-5686){\makebox(0,0)[lb]{\smash{{\SetFigFont{7}{8.4}{\familydefault}{\mddefault}{\updefault}{\color[rgb]{0,0,0}$A_2$}%
}}}}
\put(3301,-4936){\makebox(0,0)[lb]{\smash{{\SetFigFont{7}{8.4}{\familydefault}{\mddefault}{\updefault}{\color[rgb]{0,0,0}$A_1=\tilde{B}$}%
}}}}
\put(1351,-6811){\makebox(0,0)[lb]{\smash{{\SetFigFont{7}{8.4}{\familydefault}{\mddefault}{\updefault}{\color[rgb]{0,0,0}quantum system}%
}}}}
\put(151,-4936){\makebox(0,0)[lb]{\smash{{\SetFigFont{7}{8.4}{\familydefault}{\mddefault}{\updefault}{\color[rgb]{0,0,0}$B$}%
}}}}
\put(5176,-4936){\makebox(0,0)[lb]{\smash{{\SetFigFont{7}{8.4}{\familydefault}{\mddefault}{\updefault}{\color[rgb]{0,0,0}$L_{c1}$}%
}}}}
\put(5176,-6136){\makebox(0,0)[lb]{\smash{{\SetFigFont{7}{8.4}{\familydefault}{\mddefault}{\updefault}{\color[rgb]{0,0,0}$L_{c2}$}%
}}}}
\put(1801,-7186){\makebox(0,0)[lb]{\smash{{\SetFigFont{7}{8.4}{\familydefault}{\mddefault}{\updefault}{\color[rgb]{0,0,0}$\mathbf{G}_q$}%
}}}}
\put(5251,-7186){\makebox(0,0)[lb]{\smash{{\SetFigFont{7}{8.4}{\familydefault}{\mddefault}{\updefault}{\color[rgb]{0,0,0}$\mathbf{G}_c$}%
}}}}
\put(1801,-5461){\makebox(0,0)[lb]{\smash{{\SetFigFont{7}{8.4}{\familydefault}{\mddefault}{\updefault}{\color[rgb]{0,0,0}$L_{q}$}%
}}}}
\put(6301,-5311){\makebox(0,0)[lb]{\smash{{\SetFigFont{7}{8.4}{\familydefault}{\mddefault}{\updefault}{\color[rgb]{0,0,0}$\tilde{A}_1$}%
}}}}
\put(6301,-6061){\makebox(0,0)[lb]{\smash{{\SetFigFont{7}{8.4}{\familydefault}{\mddefault}{\updefault}{\color[rgb]{0,0,0}$\tilde{A}_2$}%
}}}}
\put(7201,-5911){\makebox(0,0)[lb]{\smash{{\SetFigFont{7}{8.4}{\familydefault}{\mddefault}{\updefault}{\color[rgb]{0,0,0}HD}%
}}}}
{\color[rgb]{0,0,0}\put(5401,-5086){\circle*{150}}
}%
\end{picture}%

\caption{Representation of the realistic detection scheme of  Figure \ref{fig:realistic-1} as a reducible network.}
\label{fig:realistic-1a}
\end{center}
\end{figure}
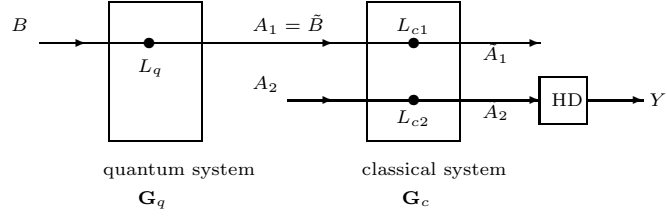

The quantum system is given by $\mathbf{G}_q= (1, L_q, H_q)$,
 and the classical detection system is given by the classical stochastic equations
 \begin{eqnarray}
d x(t) & = & \tilde f(x(t)) dt + g(x(t)) dw(t),
\nonumber
\\
d Y(t) &=& h(x(t)) dt + dv(t),
\label{system-classical-1}
\end{eqnarray}
where $x(t)\in \mathbb{R}^n$, $y(t)\in \mathbb{R}$, $\tilde f, g $
are smooth vector fields,  $h$ is a smooth real-valued function,
and $w$ and $v$ are independent standard classical Wiener
processes. As described in the Appendix \ref{sec:app-c-as-q}, this
classical system is equivalent to a commutative subsystem of
 $
\mathbf{G}_c= (1, L_{c1}, H_c)  \boxplus  (1, L_{c2}, 0)$,
 where $L_{c1}=-i g^T p-\frac{1}{2} \nabla^T g$, $L_{c2}=\frac{1}{2}h$ and $H_c=\frac{1}{2}(f^Tp+p^Tf)$.
  We represent the system of Figure \ref{fig:realistic-1} as a redicible network, as shown in  Figure \ref{fig:realistic-1a}.

 Here, the classical noises are represented as real quadratures $w = A_1 + A_1^\ast$,   $v=A_2+A_2^\ast$.
 Note that since $L_{c1}$ is skew-symmetric, only the real quadrature $w=A_1+A_1^\ast=\tilde{B}+\tilde{B}^\ast$ affects the
classical system (this captures the ideal homodyne detection). The
complete cascade system is
  \begin{eqnarray}
\mathbf{G} &=&  \left( (1, L_{c1}, H_c) \triangleleft (1, L_q, H_q)  \right) \boxplus  (1, L_{c2}, 0)
\label{realistic-2} \\
&=& (\mathbf{I}, \left(  \begin{array}{c}  L_1+L_{c1} \\ L_{c2} \end{array} \right), H_q+H_c+ \frac{1}{2i} ( L_{c1}^\ast L_q - L_q^\ast L_{c1}      )  )
\notag
\end{eqnarray}

Applying quantum filtering  \cite{VPB92a}, \cite{BHJ07},  the
unnormalized quantum filter for the cascade system $\mathbf{G}$ is
\begin{eqnarray}
 d \sigma_t(X) = \sigma_t ( -i[X, H_q+H_c+ \frac{1}{2i} ( L_{c1}^\ast L_q - L_q^\ast L_{c1}      ) ]
\notag  \\
 +\mathcal{L}_{\left(  \begin{array}{c}  L_1+L_{c1} \\ L_{c2} \end{array} \right)}(X)
 ) dt + \sigma_t( L_{c2}^\ast X + X L_{c2} ) dy .
 \label{classical-system-qf-1}
 \end{eqnarray}
Here, $X$ is any operator defined on the quantum-classical cascade
system. For instance, $X=X_q \otimes \varphi$, where $\varphi$ is a
smooth real valued function on $\mathbb{R}^n$. In particular, if
$X=X_q$ is a quantum system operator,   one can compute the desired estimate of $X_q$ from
$\pi_t(X_q)=\sigma_t(X_q)/\sigma_t(1)$. 

Equation (\ref{classical-system-qf-1}) can be normalized, and
compared with  \cite[eq. (17)]{WWM02}. In the case that the
quantum system is a linear gaussian system, and the filter is a
linear system, the complete filter reduces to a Kalman filter from
which the desired quantum system variables can be estimated.

\section{Conclusion}
\label{sec:c}

In this paper we have presented algebraic tools for modeling quantum networks. The tools include a
parametric representation for open quantum systems, and the concatenation and series products.  The concatenation product allows us to
form a larger system from components, without necessarily including  connections. The series product, through the
principle of series connections (Theorem \ref{thm:series-fb}), provides a mechanism for combining systems via field
mediated  connections. We demonstrated how to model a class of quantum networks, called reducible networks, using our
theory and we illustrated our results by examining some examples from the literature.

Future work will involve further development of the network theory described here, and applying the theory to develop
control engineering tools and to applications in quantum technology, e.g. \cite{JG07}.

\appendix

\section{Time-Ordered Exponentials in the sense of Holevo}
\label{sec:TOE}

Holevo \cite{AH92} developed a parameterization of open system
dynamics that is different to the Hudson-Parthasarathy parameters
$\mathbf{G}=\left( S,L,H\right) $. Holevo's parameterization is
defined as follows. Let
\begin{equation}
K\left( t\right) =  H_{00} t + H_{01} A(t) + H_{10} A^\ast(t)  +
H_{11}\Lambda(t) , \label{H-parameters}
\end{equation}
 where $\left\{ H_{\alpha \beta
}\right\} $ consists of bounded operators with $H_{\alpha \beta
}=H_{\beta \alpha }$, and the indices $\alpha, \beta$ range from 0
to 1 (here we are considering a single field channel for
simplicity). The \textit{time-ordered exponential with Holevo
generator}  $\left\{ H_{\alpha \beta }\right\} $ is the unitary
adapted process $U$ satisfying the quantum stochastic differential
equation
\begin{equation*}
dU\left( t\right) =\left( e^{-idK\left( t\right) }-1\right)
U\left( t\right)
\end{equation*}
with $U\left( 0\right) =1$, \cite{AH92}, \cite{JG04}.
  Expanding the  differential
$e^{-idK\left( t\right) }-1$ we  obtain
$$
dU\left( t\right) =\sum_{n\geq 1}\frac{\left( -i\right)
^{n}}{n!}\left( dK\right) ^{n} U(t) .
$$
Now for a system with parameters $\mathbf{G}=\left( S,L,H\right) $
we have
\begin{eqnarray*}
dU(t) = \{ (S-I) d\Lambda(t) +  L dA^\ast(t) - L^\ast dA 
\\
-(iH +
\frac{1}{2} L^\ast L) dt \} U(t) .
\end{eqnarray*}
Comparing these expressions, we find that
 \begin{eqnarray}
 S=\exp \left( -iH_{11}\right), L=\frac{ \exp\left(-iH_{11}
\right)-1 }{H_{11} } H_{1 0}, 
\notag \\
H=H_{00}-H_{0 1}  \frac{H_{11}
-\sin( H_{11 } )}{(H_{11})^2}
  H_{10}.
  \label{hp-holevo}
 \end{eqnarray}

The relationship between the generating coefficients $H_{\alpha
\beta }$ and the  parameters  $\mathbf{G}=\left( S,L,H\right)$ are
exactly as occur in the  implicit-explicit formalism of
\cite{HW94a},
 however, this formalism only
coincides with the Stratonovich-It\={o} correspondence in the case where $%
H_{11 }=0$ \cite{JG06}.

\section{Proof of Theorem \ref{thm:series-fb}}
\label{sec:app-series} There are a number of independent
derivations of the series product. For instance it can be derived
from a purely Hamiltonian formalism for quantum networks
\cite{GJ08a}, alternatively Gardiner's arguments in the Heisenberg
picture can be extended to include the scattering terms
\cite{G08}. Here we present a discretization argument for the
input/output fields based on \cite{JG04}. Rather than considering a continuous noise
source, we take a beam consisting of qubits (spin one-half
particles) with a rate of one qubit every $\tau $ seconds. A qubit
has the Hilbert space $\mathsf{H} = \mathbb{C}^2$ spanned by a
pair of orthogonal vectors $e_0$ and $e_1$. We define
raising/lowering operators $\sigma^\pm$ for each qubit by
$\sigma^+ (\alpha e_0 +\beta e_1)=\alpha e_1$ and $\sigma^-
(\alpha e_0 +\beta e_1)=\beta e_0$.  In our model of the interaction
of a qubit with a given plant, we shall assume that the
interaction is much shorter than $\tau $ so that at most one qubit
may interacting with a given plant at any instant of time. For two
plants in cascade, we shall take them to be separated so that the
time of flight of the qubits is exactly $\tau $ seconds. This is
purely for convenience and can be easily relaxed. For
definiteness, we assume that each qubit is prepared independently
in the ``ground state''  $e_0$ and we denote by $\sigma _{k}^{\pm
}$ the raising/lowering operators for the $k$th qubit: the
operators corresponding to different qubits commute, while we have
$\sigma _{k}^{-}\sigma _{k}^{+}+\sigma _{k}^{+}\sigma _{k}^{-}=1$,
$\left( \sigma _{k}^{+}\right) ^{2}=0=\left( \sigma
_{k}^{-}\right) ^{2}$. At time $t_{k}=k\tau $ $\left( k\in
\mathbb{N}\right) $, we take the most recent qubit to interact
with the first system to be the $k$th qubit, and the most recent
to interact with the second to be the $\left( k-1\right) $st
qubit.

Let us denote the value of $x>0$ rounded down to the nearest whole
number by $\left\lfloor x\right\rfloor $ and set
\begin{equation*}
\sigma _{\tau }^{\alpha \beta }\left( k\right) :=\left[ \frac{\sigma _{k}^{+}%
}{\sqrt{\tau }}\right] ^{\alpha }\left[ \frac{\sigma _{k}^{-}}{\sqrt{\tau }}%
\right] ^{\beta }
\end{equation*}
where $\alpha ,\beta $ may take the values zero and one and where $\left[ B%
\right] ^{0}=1$, $\left[ B\right] ^{1}=B$ for any operator $B$. In
the following, we shall denote by $O\left( \tau ^{n}\right) $ any
expression which is norm-convergent to zero as $\tau \rightarrow
0$ as fast as $\tau ^{n}$. The identity $\tau \sigma _{\tau
}^{\alpha 1}\left( k\right) \sigma _{\tau }^{1\beta }\left(
k\right) =\sigma _{\tau }^{\alpha \beta }\left( k\right) +O\left(
\tau \right) $ will be important in what follows and will
correspond to the discrete version of the second order It\={o}
products. For $t>0$ fixed, the processes
\begin{equation*}
A_{\tau }^{\alpha \beta }\left( t\right) :=\tau
\sum_{k=1}^{\left\lfloor t/\tau \right\rfloor }\sigma _{\tau
}^{\alpha \beta }\left( k\right)
\end{equation*}
are well-known approximations to the fundamental processes
$A^{\alpha \beta }\left( t\right) $ in the limit $\tau \rightarrow
0^{+}$, \cite{JG04}.


We shall fix bounded operators $H_{j}^{\alpha \beta }$ on the
$j$th system such that $H_{j}^{\alpha \beta \dag }=H_{j}^{\beta
\alpha }$ and set
$
\mathcal{H}_{\tau }^{\left( j\right) }\left( k\right)
=H_{j}^{\alpha \beta }\otimes \sigma _{\tau }^{\alpha \beta
}\left( k\right) .
$
We shall first recall some well known results \cite{JG04} for the
situation
where the qubits interact with only the first system (that is, set $%
H_{2}^{\alpha \beta }=0$). The discrete time evolution is
described by unitary kicks every $\tau $ seconds according to
$
U_{\tau }\left( t\right) =\mathcal{U}_{\left\lfloor t/\tau
\right\rfloor }\cdots \mathcal{U}_{2}\mathcal{U}_{1}
$
where $\mathcal{U}_{k}=\exp \left\{ -\mathsf{i}\tau
\mathcal{H}_{\tau }^{\left( 1\right) }\left( k\right) \right\} $.
Expanding the exponential yields
$
\mathcal{U}_{k}=1+\tau G_{1}^{\alpha \beta }\otimes \sigma _{\tau
}^{\alpha \beta }\left( k\right) +O\left( \tau ^{2}\right)
$
with the $G_{1}^{\alpha \beta }$ forming the coefficients of the
unitary QSDE with parameters $\mathbf{G}_{1}$ related to
$\mathbf{H}_{1}=\left\{ H_{\alpha \beta }^{\left( 1\right)
}\right\} $ as in Appendix \ref{sec:TOE}.

In the limit $\tau \rightarrow 0^{+}$, the discrete time process
$U_{\tau }\left( t\right) $ converges weakly in matrix elements to
the solution of the QSDE
\begin{equation*}
dU\left( t\right) =G_{1}^{\alpha \beta }\otimes dA^{\alpha \beta
}\left( t\right) \,U\left( t\right) .
\end{equation*}


We now turn to the case of a cascaded system. This time the
discrete time dynamics is given by
$
V_{\tau }\left( t\right) =\mathcal{V}_{\left\lfloor t/\tau
\right\rfloor }\cdots \mathcal{V}_{2}\mathcal{V}_{1}
$
where $\mathcal{V}_{k}=\exp \left\{ -\mathsf{i}\tau
\mathcal{H}_{\tau }^{\left( 1\right) }\left( k\right)
-\mathsf{i}\tau \mathcal{H}_{\tau }^{\left( 2\right) }\left(
k-1\right) \right\} $. Expanding the exponential now yields
\begin{equation*}
\mathcal{V}_{k}=1+\tau G_{1}^{\alpha \beta }\otimes \sigma _{\tau
}^{\alpha \beta }\left( k\right) +\tau G_{2}^{\alpha \beta
}\otimes \sigma _{\tau }^{\alpha \beta }\left( k-1\right) +O\left(
\tau ^{2}\right) .
\end{equation*}
with the $G_{2}^{\alpha \beta }$ forming the coefficients of the
unitary QSDE with parameters $\mathbf{G}_{2}$ related to
$\mathbf{H}_{2}$ as in Appendix \ref{sec:TOE}.

To better understand what is going on, we compute
\begin{eqnarray*}
\mathcal{V}_{k}\mathcal{V}_{k-1} = 1+\tau G_{1}^{\alpha \beta
}\otimes \sigma _{\tau }^{\alpha \beta }\left( k\right) 
\\
+\tau
\left\{ G_{2}^{\alpha \beta }+G_{1}^{\alpha \beta }+G_{2}^{\alpha
1}G_{1}^{1\beta }\right\}
\otimes \sigma _{\tau }^{\alpha \beta }\left( k-1\right) \\
 +\tau G_{2}^{\alpha \beta }\otimes \sigma _{\tau }^{\alpha \beta
}\left( k-2\right) +O\left( \tau ^{2}\right) .
\end{eqnarray*}
This may be iterated to give
\begin{eqnarray*}
\mathcal{V}_{k}\mathcal{V}_{k-1}\cdots \mathcal{V}_{l} =
\\
 1+\tau
\left\{ G_{2}^{\alpha \beta }+G_{1}^{\alpha \beta }+G_{2}^{\alpha
1}G_{1}^{1\beta }\right\} \otimes \sum_{j=l}^{k-1}\sigma _{\tau
}^{\alpha \beta }\left(
k-1\right) \\
  +\tau G_{1}^{\alpha \beta }\otimes \sigma _{\tau }^{\alpha \beta
}\left( k\right) +\tau G_{2}^{\alpha \beta }\otimes \sigma _{\tau
}^{\alpha \beta }\left( l-1\right) +O\left( \tau ^{2}\right) .
\end{eqnarray*}

Under the same mode of convergence as before, we obtain the limit
QSDE
\begin{equation*}
dV_{t}=G_{\alpha \beta }^{\left( 2\leftarrow 1\right) }\otimes
dA^{\alpha \beta }\left( t\right) \,V\left( t\right)
\end{equation*}
where we recognize $G_{\left( 2\leftarrow 1\right) }^{\alpha
\beta }=G_{2}^{\alpha \beta }+G_{1}^{\alpha \beta }+G_{2}^{\alpha
1}G_{1}^{1\beta } $ as the coefficients the unitary QSDE with
 the series product parameters
$\mathbf{G}_{2}\triangleleft \mathbf{G}_{1}$, see (\ref{series-diff}). Therefore $\mathbf{G}%
_{2\leftarrow 1} \equiv \mathbf{G}_{2}\triangleleft
\mathbf{G}_{1}$. The generalization to multi-dimensional noise is
straightforward.

 \section{Proof of Theorem \ref{thm:exchange}}
\label{sec:app-swap}

Clearly, if (\ref{exchange-1}) is satisfied, then both cascade
systems are described by the same parameters, which implies that
they are equivalent.
Now suppose the two systems are parametrically equivalent, with $\mathbf{S}_2'$ undetermined.
Now by Definition \ref{dfn:series} we may obtain expressions for 
$\mathbf{G}_2 \triangleleft \mathbf{G}_1$ and $ \mathbf{G}_1 \triangleleft \mathbf{G}_2'$. Equating the first terms, we have
$
\mathbf{S}_2\mathbf{S}_1 = \mathbf{S}_1\mathbf{S}_2' ,
$
and solving for $\mathbf{S}_2'$ one obtains $\mathbf{S}_{2}^{\prime } =\mathbf{S}_{1}^{\dag }\mathbf{S}_{2}\mathbf{S}_{1}$, as in (\ref{exchange-2}).
Next, equating the second terms   gives
 $
 \mathbf{L}_2+\mathbf{S}_2\mathbf{L}_1 =  \mathbf{L}_1+\mathbf{S}_1\mathbf{L}_2'.
$
This expression can be solved for $\mathbf{L}_2'$, as in  (\ref{exchange-2}).
Similarly, the Hamiltonian term $H_2'$ in  (\ref{exchange-2})  can be found by equating the third terms.

\section{Classical Systems as Commutative Quantum Subsystems}
\label{sec:app-c-as-q}

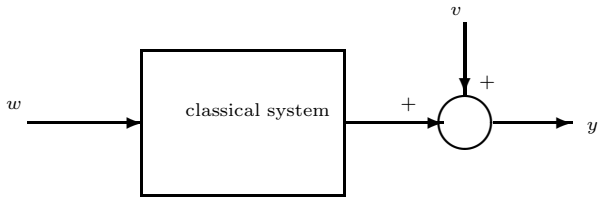
\begin{figure}[h]
\begin{center}

\setlength{\unitlength}{2368sp}%
\begingroup\makeatletter\ifx\SetFigFont\undefined%
\gdef\SetFigFont#1#2#3#4#5{%
  \reset@font\fontsize{#1}{#2pt}%
  \fontfamily{#3}\fontseries{#4}\fontshape{#5}%
  \selectfont}%
\fi\endgroup%
\begin{picture}(6351,2029)(1576,-3683)
\put(3451,-2836){\makebox(0,0)[lb]{\smash{{\SetFigFont{7}{8.4}{\familydefault}{\mddefault}{\updefault}{\color[rgb]{0,0,0}classical system}%
}}}}
\thicklines
{\color[rgb]{0,0,0}\put(3001,-3661){\framebox(2100,1500){}}
}%
{\color[rgb]{0,0,0}\put(5101,-2911){\line( 1, 0){975}}
\put(6076,-2911){\vector( 1, 0){ 75}}
}%
{\color[rgb]{0,0,0}\put(6676,-2911){\vector( 1, 0){825}}
}%
{\color[rgb]{0,0,0}\put(1801,-2911){\vector( 1, 0){1200}}
}%
{\color[rgb]{0,0,0}\put(6376,-1861){\vector( 0,-1){750}}
}%
\put(6526,-2536){\makebox(0,0)[lb]{\smash{{\SetFigFont{7}{8.4}{\familydefault}{\mddefault}{\updefault}{\color[rgb]{0,0,0}$+$}%
}}}}
\put(5701,-2761){\makebox(0,0)[lb]{\smash{{\SetFigFont{7}{8.4}{\familydefault}{\mddefault}{\updefault}{\color[rgb]{0,0,0}$+$}%
}}}}
\put(6226,-1786){\makebox(0,0)[lb]{\smash{{\SetFigFont{7}{8.4}{\familydefault}{\mddefault}{\updefault}{\color[rgb]{0,0,0}$v$}%
}}}}
\put(7651,-2986){\makebox(0,0)[lb]{\smash{{\SetFigFont{7}{8.4}{\familydefault}{\mddefault}{\updefault}{\color[rgb]{0,0,0}$y$}%
}}}}
\put(1576,-2761){\makebox(0,0)[lb]{\smash{{\SetFigFont{7}{8.4}{\familydefault}{\mddefault}{\updefault}{\color[rgb]{0,0,0}$w$}%
}}}}
{\color[rgb]{0,0,0}\put(6376,-2911){\circle{540}}
}%
\end{picture}%

\caption{Block diagram of the classical system (\ref{system-classical-1}).}
\label{fig:classical-system-1}
\end{center}
\end{figure}

In this subsection we explain how to  model the classical system
(\ref{system-classical-1}), shown in Figure
\ref{fig:classical-system-1},  as a commutative subsystem of a
larger quantum system. This representation is used in subsection
\ref{sec:eg-realistic}.  In equation (\ref{system-classical-1}),
$x(t)\in \mathbb{R}^n$, $y(t)\in \mathbb{R}$,  $\tilde f, g $ are
smooth vector fields,  $h$ is a smooth real-valued function,  and
$w$ and $v$ are independent standard classical  Wiener processes.

 To model this classical system, we take the underlying Hilbert space of the
system to be $\frak{h}=L_{2}\left( \mathbb{R}^{n}\right) $ with $q^{j}$, $%
p_{j}$ being the usual canonical position and momentum observables: $%
q^{j}\psi \left(x \right) =X_{j}\psi \left( x \right) $ and $%
p_{j}\psi \left( x \right) =-i\partial _{j}\psi \left(
\vec{x}\right)$. We write $q=(q^1,\ldots,q^n)^T$,
$p=(p_1,\ldots,p_n)^T$, and $\nabla = (\partial
_{1},\ldots,\partial _{n})^T$. If $\varphi   $ is a smooth
function of $x$, then we find that, by It\={o}'s rule, for
$\varphi_t=\varphi(x(t))$,
\begin{equation}
d \varphi = \frak{L}_{\text{classical}}\left( \varphi \right) dt + g^T \nabla \varphi \, dw ,
\label{ap-c-as-q-1}
\end{equation}
where
$
\frak{L}_{\text{classical}}\left( \varphi \right) =f^T \nabla \varphi +\frac{1}{2}g^T  \nabla \left( g^T \nabla \varphi
\right)
$
 is the (classical) generator of the diffusion process $x(t)$ in   (\ref{system-classical-1}).

\begin{figure}[h]
\begin{center}

\setlength{\unitlength}{2368sp}%
\begingroup\makeatletter\ifx\SetFigFont\undefined%
\gdef\SetFigFont#1#2#3#4#5{%
  \reset@font\fontsize{#1}{#2pt}%
  \fontfamily{#3}\fontseries{#4}\fontshape{#5}%
  \selectfont}%
\fi\endgroup%
\begin{picture}(5826,2630)(3001,-7169)
\put(5251,-5386){\makebox(0,0)[lb]{\smash{{\SetFigFont{7}{8.4}{\familydefault}{\mddefault}{\updefault}{\color[rgb]{0,0,0}$L_{c1}$}%
}}}}
{\color[rgb]{0,0,0}\thicklines
\put(5401,-5836){\circle*{150}}
}%
{\color[rgb]{0,0,0}\put(4801,-6361){\framebox(1200,1800){}}
}%
{\color[rgb]{0,0,0}\put(4276,-5086){\vector( 1, 0){2775}}
}%
{\color[rgb]{0,0,0}\put(3751,-5086){\vector( 1, 0){600}}
}%
{\color[rgb]{0,0,0}\put(4276,-5836){\vector( 1, 0){2775}}
}%
{\color[rgb]{0,0,0}\put(3751,-5836){\vector( 1, 0){600}}
}%
{\color[rgb]{0,0,0}\put(7051,-6136){\framebox(600,600){}}
}%
{\color[rgb]{0,0,0}\put(7651,-5836){\vector( 1, 0){825}}
}%
\put(4726,-6811){\makebox(0,0)[lb]{\smash{{\SetFigFont{7}{8.4}{\familydefault}{\mddefault}{\updefault}{\color[rgb]{0,0,0}classical system}%
}}}}
\put(3001,-5911){\makebox(0,0)[lb]{\smash{{\SetFigFont{7}{8.4}{\familydefault}{\mddefault}{\updefault}{\color[rgb]{0,0,0}$A_2$}%
}}}}
\put(3001,-5161){\makebox(0,0)[lb]{\smash{{\SetFigFont{7}{8.4}{\familydefault}{\mddefault}{\updefault}{\color[rgb]{0,0,0}$A_1$}%
}}}}
\put(5176,-7111){\makebox(0,0)[lb]{\smash{{\SetFigFont{7}{8.4}{\familydefault}{\mddefault}{\updefault}{\color[rgb]{0,0,0}$\mathbf{G}_c$}%
}}}}
\put(7201,-5911){\makebox(0,0)[lb]{\smash{{\SetFigFont{7}{8.4}{\familydefault}{\mddefault}{\updefault}{\color[rgb]{0,0,0}HD}%
}}}}
\put(8551,-5911){\makebox(0,0)[lb]{\smash{{\SetFigFont{7}{8.4}{\familydefault}{\mddefault}{\updefault}{\color[rgb]{0,0,0}$y$}%
}}}}
\put(6226,-6136){\makebox(0,0)[lb]{\smash{{\SetFigFont{7}{8.4}{\familydefault}{\mddefault}{\updefault}{\color[rgb]{0,0,0}$\tilde{A}_2$}%
}}}}
\put(6226,-5311){\makebox(0,0)[lb]{\smash{{\SetFigFont{7}{8.4}{\familydefault}{\mddefault}{\updefault}{\color[rgb]{0,0,0}$\tilde{A}_1$}%
}}}}
\put(5251,-6136){\makebox(0,0)[lb]{\smash{{\SetFigFont{7}{8.4}{\familydefault}{\mddefault}{\updefault}{\color[rgb]{0,0,0}$L_{c2}$}%
}}}}
{\color[rgb]{0,0,0}\put(5401,-5086){\circle*{150}}
}%
\end{picture}%

\caption{Network representation of the classical system (\ref{system-classical-1}) shown in Figure \ref{fig:classical-system-1}.}
\label{fig:classical-system-1a}
\end{center}
\end{figure}
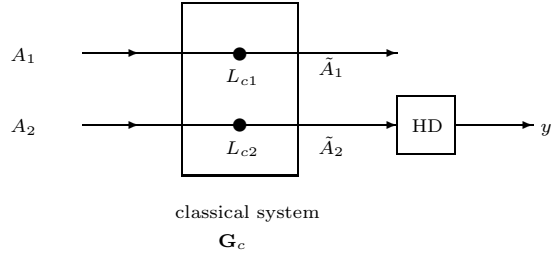

We seek a quantum network representation $\mathbf{G}_c$, as shown in Figure \ref{fig:classical-system-1a}.

The classical noises are viewed as real quadratures of quantum noises
$
w=A_1+A_1^\ast, \ \ v=A_2+A_2^\ast .
$
Now define
  port operators $L_{c1}= -i g^Tp - \frac{1}{2} \nabla^T g$, $L_{c2}=\frac{1}{2}h$ and internal Hamiltonian
  $H_c=\frac{1}{2}\left(  f^T p+p^T f\right)$,
where $f = \tilde f - \frac{1}{2} [\nabla g] g$ (the Stratonovich drift) and
$g$  are $n$-vectors whose components are viewed as  functions of
$q$ and $h=h\left( q\right) $ is viewed as a self-adjoint observable function
of $q$.
We claim that the classical system (\ref{system-classical-1}) behaves as an invariant commutative subsystem
of the open quantum system $\mathbf{G}_c= (1, L_{c1}, H_c)  \boxplus  (1, L_{c2}, 0)$.
To verify this assertion, we examine the dynamics.
From (\ref{X-qsde}) we have
\begin{eqnarray}
 dX_c = (-i[X_c,H_c] + \mathcal{L}_{L_{c1}}(X_c) + \mathcal{L}_{L_{c2}}(X_c) )dt
\notag \\
  + [X_c, L_{c1}] ( dA_1^\ast + dA_1) + [X_c, L_{c2}] (  dA_2^\ast - d A_2 )
 \label{classical-system-3}
\end{eqnarray}
Now set $X_c =\varphi = \varphi(q)$, a smooth function of the position operator. Then (\ref{classical-system-3}) gives
\begin{eqnarray}
 d \varphi &=&  (-i[\varphi,H_c] + \mathcal{L}_{L_{c1}}(\varphi) + \mathcal{L}_{L_{c2}}(\varphi) )dt
 \notag \\
 && + [\varphi, L_{c1}] ( dA_1^\ast + dA_1) + [\varphi, L_{c2}] (  dA_2^\ast - d A_2 )
  \nonumber \\
  &=& ( f^T \nabla \varphi + \frac{1}{2}g^T  \nabla \left( g^T \nabla \varphi \right)) dt + g^T \nabla \varphi \, dw ,
 \label{classical-system-4}
\end{eqnarray}
 where, we have used $-i[\varphi, H_c]=f^T \nabla \varphi$,
$\mathcal{L}_{L_{c1}}(\varphi) =  \frac{1}{2}g^T  \nabla ( g^T
\nabla \varphi)$, $\mathcal{L}_{L_{c2}}(\varphi)=0$, $ [\varphi,
L_{c1}] = g^T \nabla \varphi$, and $[\varphi, L_{c2}]=0$. Hence
the classical dynamics (\ref{ap-c-as-q-1}) is embedded in the
dynamics of the position observable $q$ only in the quantum system
$\mathbf{G}_q$ (independent of momentum dynamics). Note that only
the real quadrature of the input field affects these dynamics, and
they are unaffected by the field $A_2$.

Next we look at the outputs. The first output is not of interest,
so we focus on the second one. The output $y(t)$ of the homodyne
detector   HD in Figure \ref{fig:classical-system-1a} is
\begin{equation}
dy = d\tilde{A}_2+ d\tilde{A}_2^\ast =  (L_{c2}+L_{c2}^\ast)dt +
dA_2+ dA_2^\ast = h dt + dv \label{classical-system-5}
\end{equation}
which agrees with   (\ref{system-classical-1}), as required.
 The unnormalized quantum filter for  $\mathbf{G}_c$ is
 \begin{eqnarray}
 d \sigma_t(X_c) &=& \sigma_t ( -i[X_c,H_c] + \mathcal{L}_{L_{c1}}(X_c) + \mathcal{L}_{L_{c2}}(X_c) ) dt 
 \notag \\
 &&+ \sigma_t( L_{c2}^\ast X_c + X_c L_{c2} ) dy .
 \label{classical-system-qf-1a}
 \end{eqnarray}
 When $X_c=\varphi$, this reduces to
 \begin{equation}
 d \sigma_t(\phi) = \sigma_t (\frak{L}_{\text{classical}}\left( \varphi \right)  ) dt + \sigma_t( h \varphi) dy ,
 \label{classical-system-qf-2a}
 \end{equation}
 which is the usual Duncan-Mortensen-Zakai equation of classical nonlinear filtering, \cite[Chapter 18]{RE82}.


\bibliographystyle{plain}

\begin{thebibliography}{10}



\bibitem{VPB92a}
V.P. Belavkin.
\newblock Quantum stochastic calculus and quantum nonlinear filtering.
\newblock {\em J. Multivariate Analysis}, 42:171--201, 1992.

\bibitem{BHJ07}
L.~Bouten, R.~Van Handel, and M.R. James.
\newblock An introduction to quantum filtering.
\newblock {\em SIAM J. Control and Optimization}, 46(6):2199--2241, 2007.



\bibitem{HJC93}
H.J. Carmichael.
\newblock Quantum trajectory theory for cascaded open systems.
\newblock {\em Phys. Rev. Lett.}, 70(15):2273--2276, 1993.

\bibitem{DJ06}
C.~D'Helon and M.R. James.
\newblock Stability, gain, and robustness in quantum feedback networks.
\newblock {\em Phys. Rev. A.}, 73:053803, 2006.

\bibitem{RE82}
R.J. Elliott.
\newblock {\em Stochastic Calculus and Applications}.
\newblock Springer Verlag, New York, 1982.

\bibitem{CWG93}
C.W. Gardiner.
\newblock Driving a quantum system with the output field from another driven
  quantum system.
\newblock {\em Phys. Rev. Lett.}, 70(15):2269--2272, 1993.

\bibitem{GC85}
C.W. Gardiner and M.J. Collett.
\newblock Input and output in damped quantum systems: Quantum stochastic
  differential equations and the master equation.
\newblock {\em Phys. Rev. A}, 31(6):3761--3774, 1985.

\bibitem{GZ00}
C.W. Gardiner and P.~Zoller.
\newblock {\em Quantum Noise}.
\newblock Springer, Berlin, 2000.

\bibitem{JG04}
J.~Gough.
\newblock {H}olevo-ordering and the continuous-time limit for open {F}loquet
  dynamics.
\newblock {\em Letters in Math. Physics}, 67:207--221, 2004.


\bibitem{JG06}
J.~Gough.
\newblock Quantum Stratonovich calculus and the quantum Wong-Zakai
theorem.
\newblock {\em Journ. Math. Phys.}, 47, 113509, 2006.


\bibitem{GJ08a}
J.~Gough and M.R.~James,
\newblock Quantum Feedback Networks: Hamiltonian Formulation.
\newblock {\em Communications in Mathematical Physics}, to appear, DOI 10.1007/s00220-008-0698-8,
quant-ph/0804.3442, 2008.

\bibitem{G08}
J. ~Gough, Construction of bilinear control Hamiltonians using the
series product, {\em Phys. Rev. A}, 78, 052311, 2008.

\bibitem{GGY08}
J. ~Gough, R. Gohm, M. Yanagisawa, Linear quantum feedback
networks,  {\em Phys.Rev. A}, 78, 062104, 2008.


\bibitem{AH92}
A.S. Holevo.
\newblock Time-ordered exponentials in quantum stochastic calculus.
\newblock In {\em Quantum Probability and Related Topics}, volume~7, pages
  175--202. World Scientific, 1992.

\bibitem{HP84}
R.L. Hudson and K.R. Parthasarathy.
\newblock Quantum {I}to's formula and stochastic evolutions.
\newblock {\em Commun. Math. Phys.}, 93:301--323, 1984.

\bibitem{JG07}
M.R. James and J.~Gough.
\newblock Quantum dissipative systems and feedback control design by
  interconnection.
\newblock arxiv.org/quant-ph/0707.1074 2007.


\bibitem{JNP07}
M.R. James, H.~Nurdin, and I.R. Petersen.
\newblock ${H}^\infty$ control of linear quantum systems.
\newblock {\em IEEE Trans Auto. Control}, 53(8), 1787-1803, 2008.

\bibitem{SL00}
S.~Lloyd.
\newblock Coherent quantum feedback.
\newblock {\em Phys. Rev. A}, 62:022108, 2000.

\bibitem{EM98}
E.~Merzbacher.
\newblock {\em Quantum Mechanics}.
\newblock Wiley, New York, third edition, 1998.

\bibitem{NC00}
M.A. Nielsen and I.L. Chuang.
\newblock {\em Quantum Computation and Quantum Information}.
\newblock Cambridge University Press, Cambridge, 2000.

\bibitem{KRP92}
K.R. Parthasarathy.
\newblock {\em An Introduction to Quantum Stochastic Calculus}.
\newblock Birkhauser, Berlin, 1992.

\bibitem{HSM05}
R.~van Handel, J.~Stockton, and H.~Mabuchi.
\newblock Feedback control of quantum state reduction.
\newblock {\em IEEE Trans. Automatic Control}, 50:768--780, 2005.

\bibitem{WWM02}
P.~Warszawski, H.M. Wiseman, and H.~Mabuchi.
\newblock Quantum trajectories for realistic detection.
\newblock {\em Phys. Rev. A}, 65:023802, 2002.

\bibitem{HW94a}
H.~Wiseman.
\newblock Quantum theory of continuous feedback.
\newblock {\em Phys. Rev. A}, 49(3):2133--2150, 1994.

\bibitem{WM94b}
H.~M. Wiseman and G.~J. Milburn.
\newblock All-optical versus electro-optical quantum-limited feedback.
\newblock {\em Phys. Rev. A}, 49(5):4110--4125, 1994.

\bibitem{YK03a}
M.~Yanagisawa and H.~Kimura.
\newblock Transfer function approach to quantum control-part {I}: Dynamics of
  quantum feedback systems.
\newblock {\em IEEE Trans. Automatic Control}, (48):2107--2120, 2003.

\bibitem{YD84}
B.~Yurke and J.S. Denker.
\newblock Quantum network theory.
\newblock {\em Phys. Rev. A}, 29(3):1419--1437, 1984.


\end{thebibliography}

\end{document}